\documentstyle[multicol,pre,aps,epsf]{revtex}

\begin{document}   

\draft				     

\title{Structure of Complex-Periodic and Chaotic Media with Spiral Waves}
\author{Andrei Goryachev and Raymond Kapral\\
Chemical Physics Theory Group, Department of
Chemistry, \\ University of Toronto, Toronto, ON M5S 3H6, Canada}

\maketitle
\begin{abstract}		  
The  spatiotemporal  structure  of reactive media supporting a
solitary  spiral  wave  is studied for systems where the local
reaction  law  exhibits  a  period-doubling  cascade to chaos.
This  structure  is  considerably  more  complex  than that of
simple  period-1  oscillatory  media.  As  one  moves from the
core  of  the spiral wave the local dynamics takes the form of
perturbed,  period-doubled  orbits whose character varies with
spatial  location  relative  to the core. An important feature
of  these  media  is  the existence of a curve where the local
dynamics  is  effectively  period-1.  This  curve  arises as a
consequence   of  the  necessity  to  reconcile  the  conflict
between  the  global  topological  organization  of the medium
induced  by  the presence of a spiral wave and the topological
phase  space  structure  of  local  orbits  determined  by the
reaction  rate  law.  Due to their general topological nature,
the  phenomena  described here should be observable in a broad
class of systems with complex periodic behavior.
\end{abstract}


\begin{multicols}{2} 

\narrowtext 

\section{Introduction}\label{intro}

Spiral  waves  are spatio-temporal patterns typically found in
distributed   media  with  active  elements.  They  have  been
studied  extensively  for  excitable  and  oscillatory  media.
\cite{gen,act}  For  both  types  of media, it is conventional
to    consider   systems   with   two   dynamical   variables.
Activator-inhibitor   or   propagator-controller  systems  are
often  used  to  analyse  spiral  dynamics  in excitable media
\cite{act,fife},  while  the complex Ginzburg--Landau equation
is  the  prototypical  model  describing spatially-distributed
oscillatory media near the Hopf bifurcation point \cite{kur}.

Spiral   waves  may  also  exist  in  media  where  the  local
dynamics  supports  complex  periodic  or  even chaotic motion
that  cannot  be represented in a two-dimensional phase plane.
Various  patterns  involving  rotating  spiral waves have been
observed   in   coupled  map  lattices  or  reaction-diffusion
dynamics   based   on   the   R\"{o}ssler   chaotic  attractor
\cite{ma}.  The  three-variable reaction-diffusion system with
chaotic    local    reaction    kinetics    given    by    the
Willamowski-R\"{o}ssler  rate  law  \cite{wr} has been studied
in  \cite{prl}.  Stable  spiral waves exist in this system and
the  nucleation  and  annihilation  of spiral pairs leading to
spiral   turbulence   have   been   observed.  The  change  of
dimensionality    of   phase   space   from   two   to   three
significantly  complicates  the  description  of the dynamics.
Descriptions   in   terms   of   phase   and  amplitude,  well
established   for  two-variable  models,  cannot  be  directly
generalized.  Although  several definitions have been proposed
for  the  phase  of  chaotic  oscillations, all of them suffer
from   some   degree   of  ambiguity  (see  \cite{pik}  for  a
discussion).  Similar  difficulties arise in the consideration
of  nonchaotic  oscillatory  dynamics  which  is  nevertheless
more  complex  than a single loop in phase space; for example,
in  the  oscillations  that  appear  in  the  period  doubling
cascade  to  chaos  or in the mixed-mode oscillations observed
in experiments in chemical systems. \cite{mix}

In  this  paper  we study the spatiotemporal organization of a
reacting  medium  which  supports  a  single  spiral  wave and
where  the  local  rate  law  exhibits period-$2^n$ or chaotic
oscillations.   Through   an   analysis  of  the  dynamics  at
different  spatial  points in the medium we show that a number
of   phenomena  arise  for  $n>0$  which  are  nonexistent  in
period-1  oscillatory  media.  Section~\ref{spiral} introduces
the  model  and  presents  some  features  of  the spiral wave
behavior  in  a  chaotic  medium.  The  local  dynamics in the
medium  is  considered  in  detail  in  Sec.~\ref{local}.  The
analysis  allows  one  to  identify  the loop exchange process
for  local  trajectories  and  the  complicated pattern of the
distribution  of  different  types  of  local  dynamics in the
medium.  A  characteristic feature of this distribution is the
existence  of  a curve where the local dynamics is effectively
period-1.   Section~\ref{topo}   introduces  a  coarse-grained
description  of  $2^n$-periodic  local orbits which allows one
to  characterize  the  local  dynamics that is observed in the
medium.  The  topological  conflict  between  the  phase space
structure  of  local  trajectories and the constraints imposed
on   the   medium  by  the  existence  of  a  spiral  wave  is
considered  in  Sec.~\ref{glob}.  We  show  that  the observed
changes  of  the  local  orbits  are necessary to maintain the
global  coherence  of the medium. The conclusions of the study
are presented in Sec.~\ref{conc}.
				    
\section{Spiral Waves in Periodic and Chaotic Media}\label{spiral}

While  many  aspects  of  the  phenomena  we  describe in this
paper  are  general  and  apply  to  systems  in which complex
periodic  or  chaotic  orbits  exit,  we  consider  situations
where   a   chaotic  attractor  arises  by  a  period-doubling
cascade     and     confine    our    simulations    to    the
Willamowski-R\"{o}ssler (WR) model \cite{wr},
\begin{eqnarray}
A_1 +X &\mathrel{
\mathop{\kern0pt {\rightleftharpoons}}\limits^{{k_1}}_{k_{-1}}}& 2X,\;\;
X+Y
\mathrel{\mathop{\kern0pt {\rightleftharpoons}}\limits^{{k_2}}_{k_{-2}}}
2Y,\nonumber \\
A_5 +Y &
\mathrel{\mathop{\kern0pt {\rightleftharpoons}}\limits^{{k_3}}_{k_{-3}}}&
A_2,\;\;
X+Z
\mathrel{\mathop{\kern0pt {\rightleftharpoons}}\limits^{{k_4}}_{k_{-4}}}
A_3, \label{eq_mechanism} \\
A_4 +Z &
\mathrel{\mathop{\kern0pt {\rightleftharpoons}}\limits^{{k_5}}_{k_{-5}}}&
2Z\;. \nonumber
\end{eqnarray}
Only  the  $X$, $Y$ and $Z$ species vary with time; all others
are  assumed  fixed  by flows of reagents. Study of this model
allows  us  to  illustrate most features of the structure of a
spatially  distributed  medium  supporting  spiral  waves.  In
addition,  it  is useful to deal with a specific example since
certain  aspects  of  the  analysis  of  periodic  and chaotic
orbits  in  high-dimensional  concentration  phase spaces rely
on  geometrical  constructions  that  pertain  to  a  specific
class of attractors.

The    rate    law    that    follows   from   the   mechanism
(\ref{eq_mechanism}) is
\begin{eqnarray}
\label{mass}
{ d c_x(t) \over d t} &=&\kappa_1   c_x   -\kappa_{-1}   
c_x^2   -\kappa_2   c_x   c_y  
+\kappa_{-2} c_y^2  -\kappa_4 c_x c_z \nonumber \\ & & +\kappa_{-4} 
=R_x({\bf c}(t))\;, \nonumber \\     
{ d c_y(t) \over d t} &=&\kappa_2c_x   c_y   
-\kappa_{-2}  c_y^2   -\kappa_{3} c_y + \kappa_{-3} 
\\ & & =R_y({\bf c}(t))\;, \nonumber \\
{ d c_z(t) \over d t} &=&-\kappa_4 c_x c_z +
\kappa_{-4} +\kappa_5 c_z   -\kappa_{-5} c_z^2 
=R_y({\bf c}(t))\;, \nonumber
\end{eqnarray}
where   the   rate   coefficients   $\kappa_i$   include   the
concentrations  of  any  species held fixed by constraints. We
take  $\kappa_2$  to  be  the  bifurcation parameter while all
other     coefficients     are     fixed:     ($\kappa_1=31.2,
\kappa_{-1}=0.2,       \kappa_{-2}=0.1,       \kappa_{3}=10.8,
\kappa_{-3}=0.12,      \kappa_{4}=1.02,      \kappa_{-4}=0.01,
\kappa_{5}=16.5,  \kappa_{-5}=0.5$).  In this parameter region
the  WR  model  has  been shown \cite{ka} to possess a chaotic
attractor   arising   from   a   period-doubling   cascade  as
$\kappa_2$ is varied in the interval [1.251,1.699].

\begin{figure}[htbp]
\begin{center}
\leavevmode
\epsffile{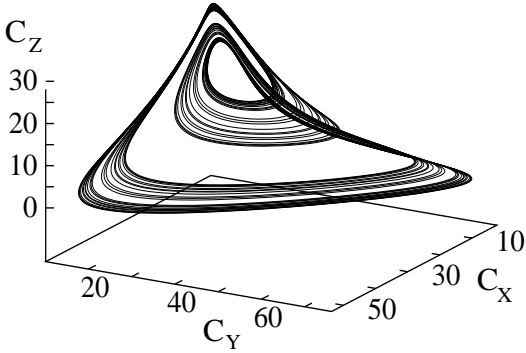}
\end{center}
\caption{Chaotic  attractor  for  the  Willamowski-R\"{o}ssler
model at $\kappa_2=1.567$. }
\label{cha}
\end{figure}

\begin{figure}[htbp]
\begin{center}
\leavevmode
\epsffile{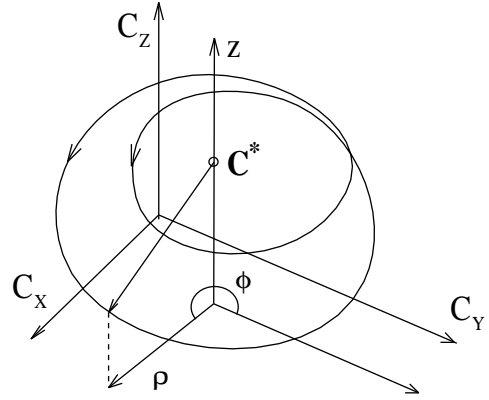}
\end{center}
\caption{Cylindrical  coordinate  frame  $(\rho,\phi,z)$  with
origin  at  ${\bf  c}^*$ in the $(c_x,c_y,c_z)$ phase space. A
period-2 orbit is shown in this coordinate frame.}
\label{frame}
\end{figure}

Figure~\ref{cha}  shows  the  four-banded chaotic attractor at
$\kappa_2  =  1.567$.  Throughout  the entire parameter domain
$\kappa_2\in   [1.251,1.699]$   the   system's   attractor  is
oriented  so  that  its  projection onto the $(c_x,c_y)$ plane
exhibits  a  folded  phase  space  flow circulating around the
unstable  focus  ${\bf  c^*}$.  This allows one to introduce a
coordinate  system  in  the  Cartesian  $(c_x,c_y,c_z)$  phase
space   which  is  appropriate  for  the  description  of  the
attractor.  We  take  the  origin  of a cylindrical coordinate
system  $(\rho,\phi,z)$  at  ${\bf  c^*}$  so that the $z$ and
zero-phase-angle   ($\phi=0$)  axes  are  directed  along  the
$c_z$  and  $c_y$  axes,  respectively. The phase angle $\phi$
increases   along   the   direction   of   flow  as  shown  in
Fig.~\ref{frame}.

For  a  period-1  oscillation  $\phi$ coincides with the usual
definition   of   the  phase  and  uniquely  parametrizes  the
attractor    $\rho_a=\rho_a(\phi),    z_a=z_a(\phi),   \phi\in
[0,2\pi)$.    After    the    first    period-doubling    this
parametrization  is  no longer unique since the periodic orbit
does  not  close on itself after $\phi$ changes by $2\pi$. For
a   period-$2^n$   orbit  $2^n$  of  its  points  lie  in  any
semi-plane   $\phi=\phi_0$.   The   angle   variable  $\Phi\in
[0,2^n\cdot   2\pi)$   may   be   used   to   parametrize  the
period-$2^n$  attractor  if  one  acknowledges that all $\Phi$
from  the  interval  $[0,2^n\cdot 2\pi)$ are different but any
two   values   of   $\Phi$,   $\Phi_1$   and   $\Phi_2$,  with
$\Phi_2=\Phi_1+   2^n\cdot   2\pi$,   are  equivalent.  For  a
chaotic  orbit  $(n  \rightarrow \infty)$ all angles $\Phi \in
[0,\infty)$  are  non-degenerate.  When  $\Phi$  is defined in
this  way  it is no longer an observable. Indeed, any $\Phi\in
[0,2^n\cdot  2\pi)$  can  be represented as $\Phi=\phi +m\cdot
2\pi$  where  $\phi\in  [0,2\pi)$  and  $m\in  {\bf N}$. While
$\phi$  is  just  the  angle coordinate in the $(\rho,\phi,z)$
system  and  is  a single-valued function of the instantaneous
concentrations      $\phi=\phi(c_x(t),c_y(t),c_z(t))$,     the
integer  number  of  turns  $m$  can be calculated only if the
entire attractor is known.

\begin{figure}[htbp]
\begin{center}
\leavevmode
\epsffile{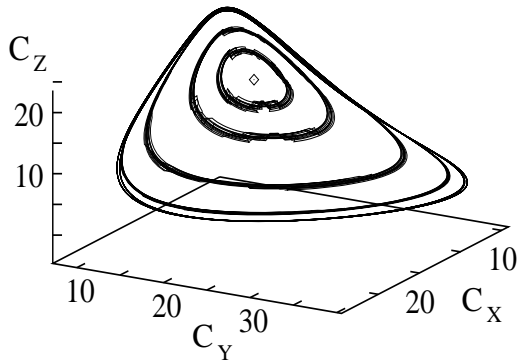}
\end{center}
\caption{Local   trajectories   calculated  for  the  period-1
oscillatory  medium  ($\kappa_2=1.420$)  at  radii  5, 10, 20,
30,  40,  56,  and  fixed  $\theta$.  The periodic orbits grow
monotonically   in  size  with  $r$;  the  difference  between
trajectories   corresponding  to  $r=40$  and  $r=56$  is  not
resolved  on  the  scale of the figure. Local orbits appear to
be  independent  on  angle  $\theta$.  The  location  of ${\bf
c}^*$ is designated by a diamond.}
\label{p-1tr}
\end{figure}

The spatially-distributed system is described by the 
reaction-diffusion equation, 
\begin{equation}
\label{rds}
{\partial {\bf c}({\bf x},t) \over \partial t} = {\bf R}({\bf c}({\bf x},t)) 
+ D \nabla^2 {\bf c}({\bf x},t)\;,
\end{equation}
where  we  have  assumed  the  diffusion  coefficients  of all
species  are  equal.  If the rate law parameters correspond to
a  period-1  limit cycle, we may initiate a spiral wave in the
medium   and   describe   its  dynamics  and  structure  using
well-developed  methods.  The  core of such a spiral wave is a
topological  defect  which is characterized by the topological
charge \cite{me}
\begin{equation} 						    
\label{charge}
{1 \over 2 \pi} \oint \nabla \phi({\bf  r})\cdot d{\bf l}=n_t\;,  
\end{equation}  
where  $\phi({\bf  r})$ is the local phase and the integral is
taken  along  a closed curve surrounding the defect. To obtain
additional   insight  into  the  organization  of  the  medium
around  the  defect  the local dynamics may be considered. For
this  purpose  we  introduce  a  polar coordinate system ${\bf
r}={\bf  x}-{\bf  r}_d(t)=(r,\theta)$  centered  at the defect
whose  (possibly  time-dependent  position) is ${\bf r}_d(t)$.
Let  ${\bf  c}({\bf r},t)$ be a vector of local concentrations
at  space  point  ${\bf  r}=(r,\theta)$. A local trajectory in
the  concentration  phase  space  from $t=t_0$ to $t=t_0+\tau$
at point ${\bf r}$ in the medium will be denoted by
\begin{equation}
C({\bf   r}|t_0,\tau)   =   \{   {\bf  c}({\bf  r},t)|  t\in
[t_0,t_0+\tau]\}\;.
\end{equation}
Figure~\ref{p-1tr}   shows  a  number  of  local  trajectories
$C(r,\theta|  t_0,\tau)$  at points with increasing separation
$r$  from  the  defect for a period-1 oscillation at $\kappa_2
=1.420$.  One  sees  that as $r \rightarrow 0$ the oscillation
amplitude  decreases  and the limit cycle shrinks to the phase
space  point  ${\bf  c}^*_d$ corresponding to the spiral core.
The  results  of  our simulations show that the value of ${\bf
c}^*_d$  differs  only  slightly  from  ${\bf  c^*}$  which is
chosen    as    the    origin    of   the   coordinate   frame
$(\rho,\phi,z)$.  Thus,  the angle $\phi$ can serve as a phase
that  characterizes  all  points  in  the period-1 oscillatory
medium  except  for  a  small  neighborhood of the defect with
radius  $r\approx  1$.  \cite{amend}  The  concentration field
${\bf  c}({\bf  r},t)$  is organized so that the instantaneous
$(c_x,c_y,c_z)$   phase  space  representation  of  the  local
concentration  on  any  closed  path in the medium surrounding
the  defect  is  a simple closed curve encircling ${\bf c^*}$.
For   large   $r$,   $r\geq   r_{max}$   (in  Fig.~\ref{p-1tr}
$r_{max}\approx  40$),  one finds that $C(r,\theta| t_0,\tau)$
ceases  to  change  shape  and  is  indistinguishable from the
period-1 attractor of (\ref{mass}) on the scale of the figure.

One  may  initiate  the  analog  of a defect in $2^n$-periodic
and  chaotic  media. The defect serves as the core of a spiral
wave  which  may  exist  even if the oscillation is not simply
period-1.  A  defect  was  introduced  in  the  center  of the
medium   by   fixing  $c_z({\bf  r})=c^{*}_{z}$  and  choosing
initial  concentrations  $  (c_x({\bf  r}),c_y({\bf  r}))$  to
produce  orthogonal  spatial  gradients.  The influence of the
symmetry   of   the   spatial   domain  on  the  dynamics  was
investigated  by  performing  simulations on square $(L \times
L)$  arrays  as  well  as  on  disk-shaped domains with radius
$R$.  No-flux  boundary  conditions  were  used to prevent the
formation  of  defects with opposite topological charge within
the   medium   and   to  minimize  effects  arising  from  the
self-interaction   of  spiral  waves.  The  implementation  of
these  initial  and boundary conditions does not guarantee the
formation  of  a solitary stable spiral wave; new spiral pairs
and  other  patterns  (e.g. pacemakers) may appear as a result
of  instabilities  of  the  spiral  arm  and  lead  to  spiral
turbulence.  The  ability to maintain a stationary spiral wave
in  the  center  of the medium is sensitive to the parameters.
For  various  values of the system size and rate constants the
defect   can   move  along  expanding  or  contracting  spiral
trajectories   or   trajectories  with  complex  ``daisy"-like
forms  \cite{daisy}.  Simulations show that the stability of a
spiral  wave  with  a stationary core located at the center of
the  medium  increases  with  the  system  size  and  for rate
constants  lying  close  to  the  chaotic  regime  within  the
period-doubling  cascade.  In  the  following  we restrict our
considerations  to  parameters that lead to the formation of a
single  spiral  wave  whose core is stationary and lies in the
center  of  the  domain.  Long  transient times ($\approx10^2$
spiral   revolutions)   are  often  necessary  to  reach  this
attracting state.

\begin{figure}[htbp]
\begin{center}
\leavevmode
\epsffile{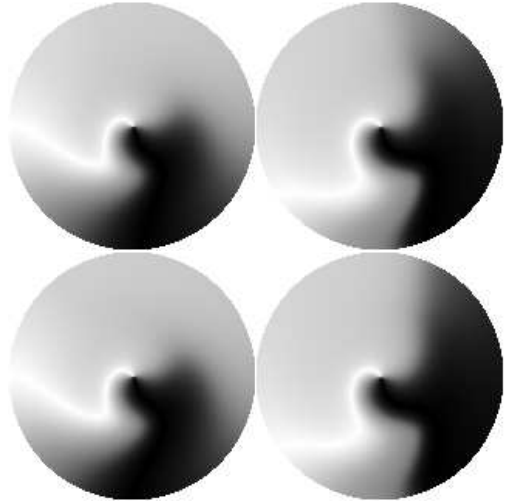}
\end{center}
\caption{Frames   showing   a  rotating  spiral  wave  in  the
chaotic  ($\kappa_2=1.567$)  disk-shaped  medium  with $R=80$.
The  local  angle variable $\phi(r,\theta,t)$ is shown as grey
shades.  Time  increases  from  left  to right and from top to
bottom.  The  frames  are  separated  by  one period of spiral
revolution  $T_r$.  The  integration  time  step is $\Delta t=
10^{-4}$  and  the  scaled  diffusion coefficient is $D \Delta
t/ (\Delta x)^2 = 10^{-2}$.}
\label{4sp}
\end{figure}
	    
Figure~\ref{4sp}   shows   four   consecutive  states  of  the
disk-shaped  medium  with  $R=80$,  separated by one period of
the  spiral  rotation,  $T_r$, for $\kappa_2 =1.567$ where the
rate   law   supports  a  chaotic  attractor.  Only  within  a
sufficiently   small   region   with  radius  $r  \approx  20$
centered  on  the  defect  does  the medium return to the same
state  after  one period of spiral rotation. At points farther
from  the  defect  the  system  appears  to return to the same
state  only  after two spiral rotation periods. The transition
from  period-1  to period-2 behavior occurs smoothly along any
ray emanating from the defect.

\section{Analysis of Local Dynamics} \label{local}
	     
\begin{figure}[htbp]
\begin{center}
\leavevmode
\epsffile{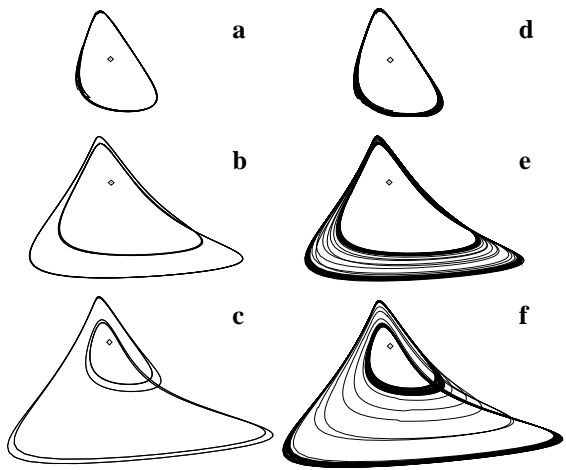}
\end{center}
\caption{Local    trajectories   $C(r|\,t_0,\tau)$   for   the
disk-shaped  medium  ($\kappa_2=1.567,\;R=80$): (a, d) $r=10$;
(b,  e)  $r=35$;  (c,  f)  $r=76$.  The  observation times are
$\tau\approx10T_r$  for  left  column  and  $\tau  =80T_r$ for
right  column.  All  the  trajectories  are  shown on the same
scale.}
\label{6traj}
\end{figure}

More   detailed   information   may   be   obtained   from  an
investigation  of  the local dynamics of the medium supporting
a  spiral  wave. Local trajectories $C({\bf r}|t_0,\tau)$ were
computed  along  rays  emanating  from  the  defect at various
angles   $\theta$.   Figure~\ref{6traj}  (left  column)  shows
short-time  trajectories  ($\tau  \approx 10T_r$) at different
radii  $r$  and  arbitrary but large $t_0$. These trajectories
clearly   demonstrate   that   the  local  dynamics  undergoes
transformation  from  small-amplitude period-1 oscillations in
the  neighborhood  of the defect to period-4 oscillations near
the boundary.\cite{f3}

The   well-resolved   period-doubling   structure  of  $C({\bf
r}|t_0,\tau)$  is  destroyed if the time of observation $\tau$
becomes    sufficiently    large.    The   right   column   of
Fig.~\ref{6traj}   shows  trajectories  sampled  at  the  same
spatial  locations  but  with  the time of observation $\tau =
80  T_r$.  These  long-time trajectories appear to be ``noisy"
period-1  and  period-2 orbits: the trajectory in panel (d) is
a  thickened  period-1  orbit  while  both the period-2 (panel
(b))  and  period-4 (panel (c)) orbits now appear as thickened
period-2   orbits  in  panels  (e)  and  (f)  with  trajectory
segments  lying  between  the  period-2 bands. As $\tau$ tends
to   infinity   the  resulting  local  attractor,  $C(r)$,  is
independent of $t_0$ and the angle $\theta$.

\subsection{First-return maps}
An   analysis   of  the  local  trajectories  shows  that  the
period-doubling  phenomenon  is  not  a  monotonic function of
$r$.   Consider  the  first  return  map  constructed  from  a
Poincar\'{e}   section   of   a   local   trajectory   $C({\bf
r}|t_0,\tau)$  in  the  following way: choose the plane $c_y =
c_y^{*}$  with  normal  ${\bf  n}$ along the $c_y$ axis as the
surface  of  section  and  select  those  intersection  points
where  ${\bf  n}$  forms  a positive angle with the flow. This
yields  a  set  $\{(c_x({\bf  r},t_n),c_z({\bf  r},t_n))| n\in
[1,N]\}$   where  $t_0  <t_1<t_2<\ldots  <t_N<t_0+\tau$  is  a
sequence   of  times  at  which  the  trajectory  crosses  the
surface  of  section.  For  the WR model the points $(c_x({\bf
r},t_n),c_z({\bf  r},t_n))$  lie  on  a  curve  which deviates
only  slightly  from  a  straight  line.  Consequently, we may
choose  either  $c_x$  or  $c_z$ to construct the first return
map.  Let  $\xi_n({\bf r})=c_x({\bf r},t_n)$ denote a point in
the   Poincar\'{e}   section.   The  relation  $\xi_{n+1}({\bf
r})=f(\xi_{n}({\bf     r}))$     between     the    successive
intersections  of  the  Poincar\'{e} surface defines the local
first return map,
\begin{eqnarray}
g({\bf  r}|t_0,\tau)  &  =  &\{ (\xi_n({\bf r}),\xi_{n+1}({\bf
r}))  \nonumber  \\  &  & |\,t_n\in [t_0,t_0+\tau], n\in [1,N]
\}\;.
\end{eqnarray}
Combining  such  maps  for  all  $r$  along some ray emanating
from   the   defect  at  an  angle  $\theta$,  we  obtain  the
cumulative first return map,
\begin{equation}
G(\theta|t_0,\tau)= \bigcup_{r\in  (0,R)}  g(r,\theta |t_0,\tau)\;.
\end{equation} 

\begin{figure}[htbp]
\begin{center}
\leavevmode
\epsffile{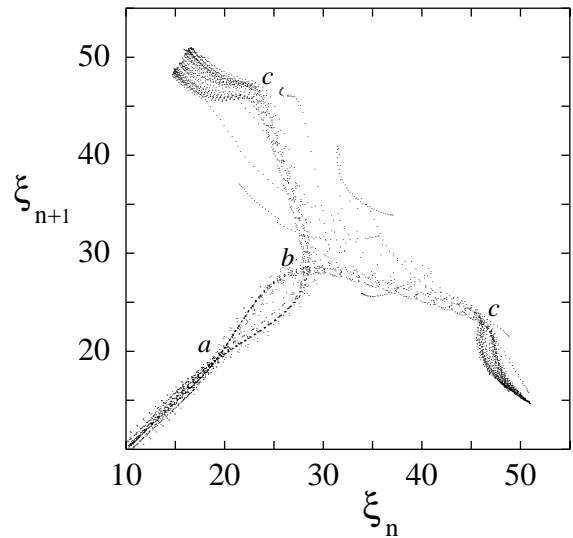}
\end{center}
\caption{Cumulative  first  return map $G$ constructed for the
disk-shaped   array   ($\kappa_2=1.567,\;R=80$).  The  letters
indicate  the  $r$  values  discussed in the text: (a) 20, (b)
31 and (c) 43.}
\label{r-frm}
\end{figure}

For  sufficiently  long  times  $\tau$,  $g$ is independent of
$\theta$   and   $t_0$.   Letting   $\lim_{\tau   \to  \infty}
g(r,\theta  |t_0,\tau)=  g(r)$, we may write the corresponding
cumulative  first  return  map  as  $G=\lim_{\tau  \to \infty}
G(\theta|t_0,\tau)$.  Figure~\ref{r-frm}  shows  $G$  for  the
disk-shaped  medium  under consideration. The first return map
is  comprised  of  several branches which can be identified as
thread-like  maxima  of  the  first  return map point density.
These  branches  are  parametrized  by  the spatial coordinate
with  $r$  increasing  from the bottom left corner to the ends
of   the  wide-spread  arms  of  $G$  (cf.  Fig.~\ref{r-frm}).
Generally  for  $r\leq  40$  points lying on lines $\xi_n(r) +
\xi_{n+1}(r)  =  {\rm const}$ belong to the same $g(r)$ though
overlaps  of  neighboring  $g$-map  points  are  common. Thus,
measuring  the  separation  between  branches  of  $G$  in the
direction  perpendicular  to  the  bisectrix one can determine
the  character  of  $C(r)$.  In spite of some evidence of fine
structure,  from  the  fact  that map points are located along
the  bisectrix  in  Fig.~\ref{r-frm}  one can infer that up to
$r=20$   the   local   dynamics   is  predominantly  period-1.
Starting  from  $r=21$  (labeled  by $a$ in Fig.~\ref{r-frm}),
$G$   splits   into   two  branches  which  diverge  from  the
bisectrix  indicating  a  period-2 structure of $C(r)$. As $r$
increases  these  branches  bend  and  cross  the bisectrix at
$r=31$  (labeled  by  $b$  in  Fig.~\ref{r-frm}), indicating a
return  of  the  local  dynamics to the period-1-like pattern.
After  this  crossing  the  separation  between  the  branches
grows   rapidly   reflecting   the   development  of  period-2
structure.   An  examination  of  the  main  branches  of  $G$
reveals  period-4  fine  structure. This period-4 structure is
visible  for  $r>28$  and beyond $r\approx 43$ (labeled by $c$
in  Fig.~\ref{r-frm})  it  becomes prominent and can be easily
seen in the structure of $C(r)$ (cf. Fig.~\ref{6traj}).

\subsection{Loop exchange and $\Omega$ curve}	

\begin{figure}[htbp]
\begin{center}
\leavevmode
\epsffile{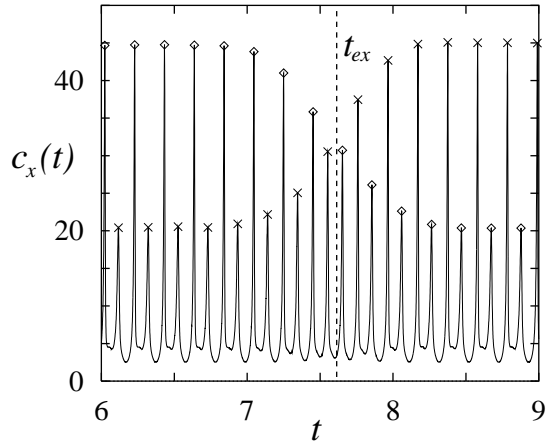}
\end{center}
\caption{Concentration  time  series  $c_x(r,t)$ at $r=50$ for
the  disk-shaped  array  ($\kappa_2=1.544,\;R=80$) showing the
loop exchange process. Time unit equals $10^5$ $\Delta t$.}
\label{p2ex}
\end{figure}

~From   the   analysis   of  the  time  series  of  the  local
concentration  one  may  determine  the  processes responsible
for  the  differences  between  the local trajectories $C({\bf
r}|t_0,\tau)$  for  short  and long time intervals $\tau$ (cf.
Fig.~\ref{6traj}).  Figure~\ref{p2ex}  shows  the signature of
this  phenomenon  for  $c_x(r,t)$  at  $r=50$ in a disk-shaped
array  with  $R=80$  and $\kappa_2 = 1.544$, a parameter value
corresponding  to  period-4  dynamics  in  the rate law. Every
second  maximum  of  $c_x(r,t)$  is  indicated  by  diamond or
cross  symbols.  The  envelope curves obtained by joining like
symbols  cross  at  $t=t_{ex}$, thus the curve which connected
large-amplitude   maxima  at  $t<t_{ex}$  joins  low-amplitude
maxima  at  $t>t_{ex}$ and vice-versa. This implies that if at
some  $t_0<t_{ex}$  the  representative point ${\bf c}(r,t_0)$
was  found  on  the  small-amplitude band of period-2, then at
$t  =t_0+nT_2>t_{ex}$,  where  $T_2$  is  the  period  of  the
period-2    oscillation,    it    will   be   found   on   the
larger-amplitude   band.\cite{f4}   This   phenomenon  can  be
interpreted  as  an  exchange  of the local attractor's bands.
Indeed,  approaching  $t_{ex}$  from  the  left one finds that
with  each  period  of  oscillation  the  small-amplitude band
grows  while  large-amplitude band shrinks. At $t=t_{ex}$ both
bands reach and pass each other.

For  a  short  period  of  time  near  $t_{ex}$  the bands are
indistinguishable  in  phase  space  and  the  oscillation  is
effectively  period-1.  It  is  this  exchange phenomenon that
produces  loops  that  fill the gap between the period-2 bands
in  the  long-time  local  trajectories (cf. Fig.~\ref{6traj})
and  contribute  to  a  sparsely scattered ``gas"-like density
in $G$ (cf. Fig.~\ref{r-frm}).

An  examination  of  the loop exchanges at different locations
in   the  medium  revealed  the  existence  of  the  following
spatio-temporal  pattern.  At  any fixed location the exchange
occurs   periodically,   with  period  $T_{ex}\approx  55T_r$,
independent  of  the  position $(r,\theta)$ in the medium. For
sufficiently  large  radii $(r\geq 35)$ this periodicity takes
an   even  stronger  form:  the  entire  oscillation  pattern,
however  complex,  returns  with  period  $T_{ex}$ to the same
configuration.
\begin{figure}[htbp]
\begin{center}
\leavevmode
\epsffile{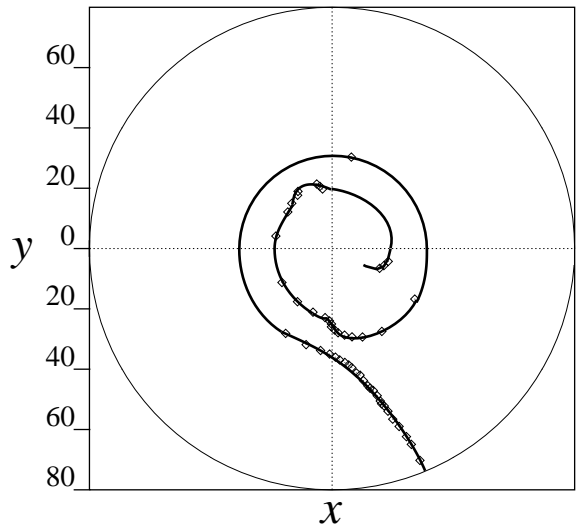}
\end{center}
\caption{Sketch  of  the  $\Omega$  curve  for the disk-shaped
array  ($\kappa_2=1.567,\;R=80$).  Points  where  the period-2
band exchange was observed are indicated by diamonds. }
\label{r-om}
\end{figure}
This   property   smoothly   disappears   as   the  defect  is
approached.  For  two locations ${\bf r}_1=(r_0,\theta_1)$ and
${\bf  r}_2=(r_0,\theta_2)$  at the same radius $r_0$ from the
defect  but  at  different  angles, the oscillation pattern at
one  of  them,  say  ${\bf  r}_2$,  can  be  obtained from the
corresponding  pattern  at  ${\bf r}_1$ through translation in
time  by  $T_{ex}(\theta_2-\theta_1)/2\pi$,  the  sign  of the
translation        being        defined        by        ${\rm
sign}(\theta_2-\theta_1)$.  In  view of this observation it is
convenient  to  introduce  a  coordinate system $(r',\theta')$
rotating  with  angular velocity $2\pi/T_{ex}$ relative to the
laboratory-fixed   coordinate  system  $(r,\theta)$.  In  this
rotating   frame   the   local  dynamics  is  described  by  a
time-homogeneous  pattern,  unique  for  every  spatial  point
${\bf  r'}$,  and  the  locations  in  the  medium  where loop
exchange   occurs   correspond   to  points  where  the  local
dynamics  always  has  a  period-1-like  character. The set of
loop  exchange  points constitute a curve $\Omega$ with spiral
symmetry   which   winds   twice   around   the   defect  (see
Fig.~\ref{r-om}).  The  two convolutions of $\Omega$ lie close
to  circular  arcs  with  radii  19 and 32. This result may be
compared  with  the  data  obtained from an examination of $G$
(cf.  Fig~\ref{r-frm}).  The  crossings  of  the  bands of $G$
occur at loci lying on $\Omega$.

Close  to  the  defect  the  resolution  of  the loop exchange
event  is  difficult.  At  $r<18$  the  difference between the
period-2  bands  is  comparable  to the band thickness and the
determination   of   $\Omega$   for   smaller   radii  becomes
impractical.  Variation  of the system parameters results in a
change  of  the  characteristics of $\Omega$; for example, the
radius  of  the  domain  $R$  does not affect the shape of the
$\Omega$  but  does change the angular velocity with which the
coordinate   frame   $(r',\theta')$   in   which  $\Omega$  is
immobile   rotates  relative  to  the  laboratory-fixed  frame
$(r,\theta)$.  The  angular  velocity  is  higher  for smaller
system  sizes:  a  decrease  in  $R$ from 80 to 60 reduces the
period  $T_{ex}$  by  a  factor  of 0.42. A change in the rate
constants  $\kappa_i$  leads  to  a  deformation  of $\Omega$,
although  the  identification of $\Omega$ as a set of exchange
points  remains  and  it  retains  the  topology  of  a  curve
passing  from  the  defect to the boundary. In Sec.~\ref{glob}
we  shall  show  that  the  existence of $\Omega$ is essential
for  the  maintenance  of spatial continuity in media composed
of $2^n$-periodic oscillators.

Simulations  on  a  square  array with dimension $80\times 80$
(all  parameters  were  the  same as for the disk) show that a
rotating    frame    is   not   necessary   to   observe   the
time-homogeneous      local      dynamics      of      $C({\bf
r}|t_0,\tau)=C({\bf  r}|\tau)$.  For  this system geometry the
$\Omega$ curve is fixed in the medium,

\end{multicols}
\widetext
\begin{figure}[htbp]
\begin{center}
\leavevmode
\epsffile{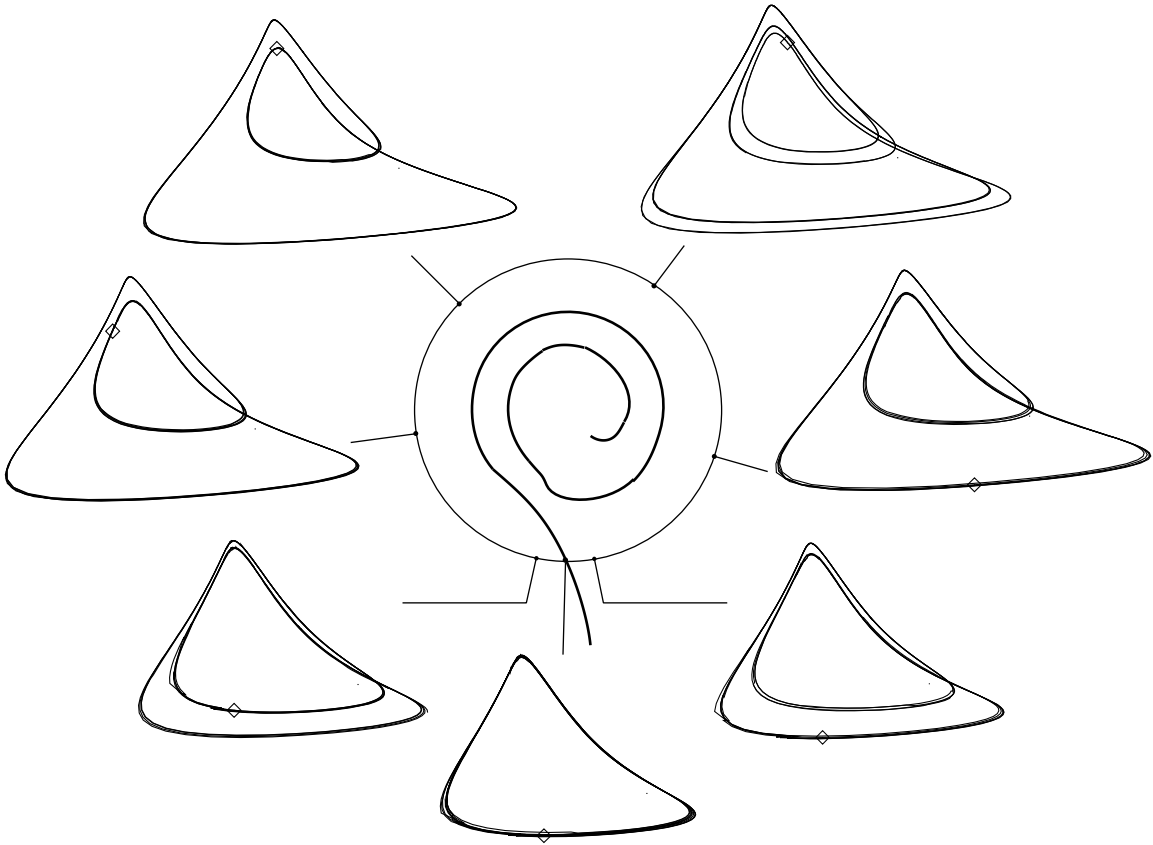}
\end{center}
\caption{Local  trajectories  calculated on circle with $r=55$
for  the  square  array. All the trajectories are shown on the
same scale.}
\label{circ}
\end{figure}
\begin{multicols}{2}
\narrowtext

\noindent  a  slight  wobbling  of  the  defect (frame origin)
being   neglected.   Figure~\ref{circ}   shows   a  number  of
long-time  $(\tau  \gg  T_r)$  local  trajectories on a circle
with  radius  $r_0=55$  surrounding  the  defect in the square
domain.  One  sees  a  significant  dependence of the shape of
$C(r_0,\theta|\tau)$   on   the   angle  $\theta$.  The  local
trajectories  range  from a period-1 orbit at the intersection
with   $\Omega$   to   the   well-established  period-4  orbit
observed  in  a  certain  range  of $\theta$. To highlight the
loop  exchange  phenomenon,  a particular time instant $t=t^*$
is  marked  on  all  the  trajectories  (see Fig.~\ref{circ}).
Compare   the   two   $C(r_0,\theta|\tau)$  at  the  locations
$\theta_1$  and  $\theta_2$  chosen  symmetrically  on  either
side  of  the  point $\theta=\theta_{\Omega}$ where the circle
intersects  $\Omega$.  Visual inspection of these orbits shows
that    their    shapes    are   essentially   identical   but
representative   points   ${\bf  c}(\theta_1,t^*)$  and  ${\bf
c}(\theta_2,t^*)$  appear  on  different period-2 bands of the
corresponding  orbits.  This  clearly  demonstrates  that  the
period-2  bands  do  not  just  approach  but indeed pass each
other    at    $\theta=\theta_{\Omega}$,    exchanging   their
positions  in  phase  space. Since it is not necessary to work
in  a  rotating  coordinate  system  in  the  case of a square
domain,  one  may  resolve  the  fine  structure  of the local
trajectories to a greater degree as can be seen
\begin{figure}[htbp]
\begin{center}
\leavevmode
\epsffile{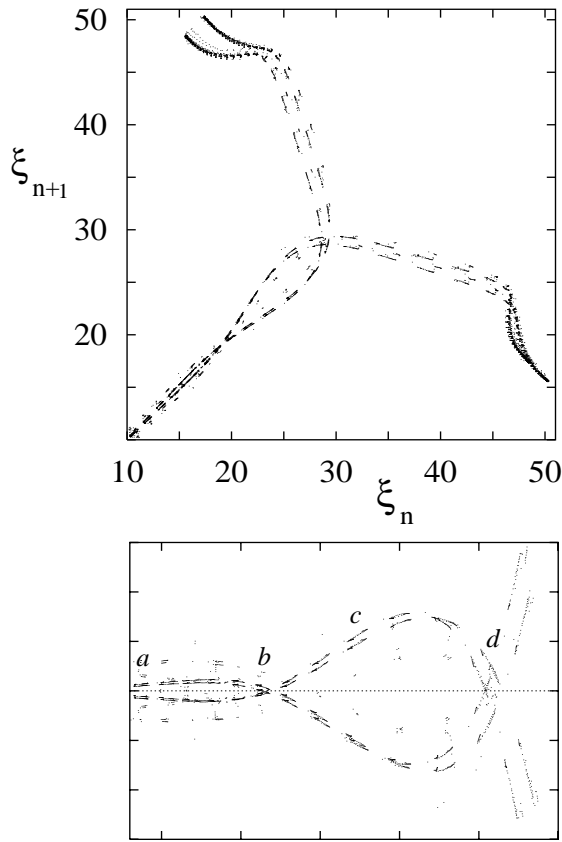}
\end{center}
\caption{Cumulative  first  return  map  $G(\theta)$  for  the
square   array   ($\kappa_2=1.567,\;L=80,\;   \theta=0$)  (top
panel)  and  a  magnification  of  a  portion of its structure
(bottom  panel).  Letters on the bottom panel denote radii for
which  corresponding  portions of $G(\theta)$ are constructed:
(a) 9; (b) 19; (c) 25, and (d) 31.}
\label{s-frm}
\end{figure}
\noindent   in   Fig.~\ref{s-frm}   (a,b)   which   shows  the
cumulative  first  return  map $G(\theta)$ and a magnification
of    a    portion    of    its    structure   (compare   with
Fig.~\ref{r-frm}).  The  results show that $G$ is comprised of
four  branches  with  the  fine structure of period-4 resolved
even  in  the  vicinity  of  the  defect ($r=5$ is the closest
distance  to  the  defect  for  which  $g(r)$  is  shown). Any
perturbation   of   the   self-organized   pattern   of  local
oscillator  synchronization  due  to  irregular  motion of the
defect,  influence  of the boundary or the presence of another
defect  may  obliterate  subtle  fine  structure  of the local
trajectories.  In  such  a circumstance one is able to observe
only  two  gross branches of $G$ and their split nature is not
resolvable  except  for  very  large  $r$.  These observations
allow  one  to  suppose  that  the local trajectories may have
the  same  number  of  fine structure levels everywhere in the
medium  but  the degree to which different levels are resolved
in  their  phase portraits strongly depends on the position in
the   medium   relative   to  the  defect.  In  view  of  this
hypothesis  the  phenomenon  of spatial period-doubling should
not  be  understood  in  the  literal  sense  but rather as an
enhanced  ability  to  resolve  the  fine  structure  with the
increase of separation from the defect.

The  stationary  rotating  spiral wave arises from the complex
defect-organized   cooperation   of  local  oscillators.  Each
location  in  the  medium  develops some site-specific pattern
of  oscillation  which  often  differs significantly from that
of   the   corresponding   rate   law   attractor  and  varies
substantially  from  one  space point to another. There exists
a   (possibly   rotating)   reference   frame  $(r',\theta')$,
centered  on  the moving defect, in which local dynamics takes
a  simple,  time-homogeneous form. Each point of the medium in
this  frame  can  be  assigned  a  unique oscillatory pattern,
different  for  different  spatial  points. This allows one to
introduce  the  notion  of a defect-organized field associated
with  $(r',\theta')$  which  specifies the pattern of dynamics
in  every  spatial  point of the medium. This field exhibits a
complicated  architecture  lacking  of  any  simple symmetries
(which  can  be  easily  seen  from  the shape of the $\Omega$
curve).  The  slow  rotation  of  this  field  in  disk-shaped
arrays   restores  the  circular  symmetry  of  the  solution.
Although   the  manner  in  which  different  types  of  local
dynamics  are  distributed in the medium is complex, it is not
disordered.  Due  to  the  continuity of the medium maintained
by  the  diffusion,  it  obeys  certain topological principles
studied in the subsequent sections.

\section{Coarse-grained description of local trajectories} \label{topo}
	    
In  the  previous  section the phase space shapes of the local
trajectories  were  shown  to  vary  considerably but smoothly
from  one  point  in  the  medium  to another. To describe the
transformations  of  these  orbits  into  each  other,  it  is
useful   to   introduce  a  description  which  captures  only
topologically  significant  changes  of  phase  portraits  and
disregards    unimportant    details.    To   understand   the
topological    principles    which    determine   the   global
organization  of  the  defect-organized field one also needs a
means  to  compare  the time dependence of local trajectories.
In  this  section  we  present  a  scheme  that  allows one to
partition   the   continuum   of   all   the   observed  local
trajectories   into   a  finite  number  of  discrete  classes
according to their phase space shape and time dependence.

\subsection{Representation of attractors by closed braids}

\begin{figure}[htbp]
\begin{center}
\leavevmode
\epsffile{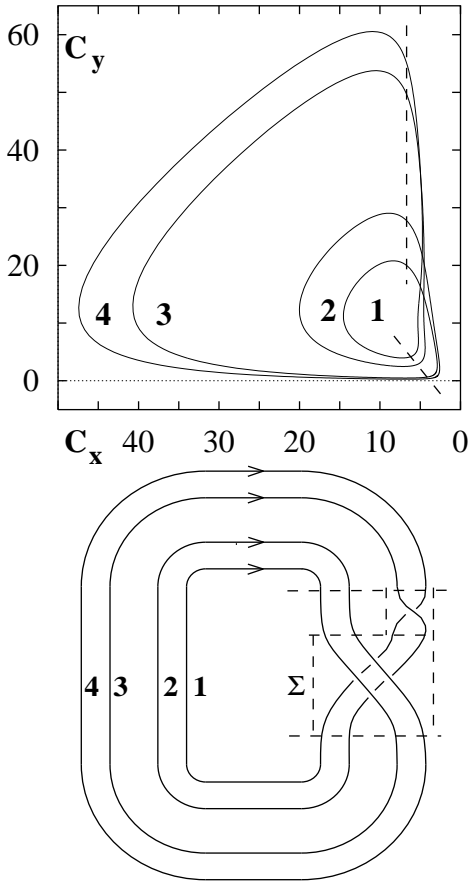}
\end{center}
\caption{Projection    of   the   $P_4$   attractor   on   the
$(c_x,c_y)$  plane  (top  panel)  and the corresponding closed
braid ${\bar B}_4$ (bottom panel). }
\label{p-4toB}
\end{figure}

Consider  a  period-$2^n$  attractor, $P_{2^n}$, consisting of
$2^n$  loops  in  the  concentration  phase  space ${\cal P} =
(c_x,c_y,c_z)$.   Using   the  cylindrical  coordinate  system
introduced   earlier,   we   may   project  $P_{2^n}$  on  the
$(\rho,\phi)$  plane  preserving  its original orientation and
3D     character     by    explicitly    indicating    whether
self-intersections  correspond  to  over  or  under crossings.
Such  a  projection shows a span of $\phi$ free from crossings
where  loops  are  essentially  parallel  to  each other. This
span  can  be used to number loops, say, in the order of their
separation  from  the  origin.  This  procedure maps $P_{2^n}$
onto    a    closed   braid   ${\bar   B}_{2^n}$   \cite{bir}.
Figure~\ref{p-4toB}   illustrates   the  construction  of  the
braid representation for the $P_4$ attractor of the WR model.

It   is  convenient  to  subdivide  the  closed  braid  ${\bar
B}_{2^n}$  into  the open braid $B_{2^n}$ (separated by dashed
lines  in  Fig.~\ref{p-4toB})  and  its  closure where threads
run  parallel  to each other. The direction of the flow on the
attractor  is  indicated  by  the arrows. Each crossing on the
projection   of  $P_4$  corresponds  to  an  elementary  braid
$\sigma_i$   which   refers   to  the  fact  that  thread  $i$
overcrosses  thread  $i+1$  (cf.  Fig.~\ref{brth}  in Appendix
for  notation  rule).  An under crossing will be designated by
${\sigma}_i^{-1}$.  A  braid  may be described by a braid word
that   gives  the  order  and  types  of  crossings  of  braid
threads.  For  example,  for the closed braid corresponding to
$P_4$  (cf.  Fig.~\ref{p-4toB})  $P_4  \mapsto  {\bar  B}_4  =
\overline{\sigma_3\sigma_2\sigma_1\sigma_3\sigma_2}$.      The
closed  braid${\bar  B}_{2^n}$  corresponding to $P_{2^n}$ can
be   represented   by   several  braid  words,  which  can  be
transformed  into  one  another by a set of allowed moves (see
Appendix).

Any  braid  word  representing $P_{2^n}$ induces a permutation
$\pi^{(n)}_i$   describing   the   order  in  which  loops  of
$P_{2^n}$   are   visited   during   one   oscillation  period
$T_{2^n}$.   In   general,   each   $P_{2^n}$   attractor   is
represented  by  several  possible $\pi^{(n)}_i$, their number
growing  with  $n$;  for  example, for $P_2$ there is only one
permutation     $\pi^{(1)}_1=    (^{12}_{21})$    while    two
permutations    $\pi^{(2)}_1    =   (^{1234}_{3421})$   (which
corresponds  to  the  braid  shown  in  Fig.~\ref{p-4toB}) and
$\pi^{(2)}_2  =  (^{1234}_{4312})$  exist  for  $P_4$.  With a
given  loop  numbering convention each braid word represents a
unique  permutation  while  one  permutation can be induced by
many braid words.

\subsection{Symbolic representation of periodic orbits}
				   
Take  two  period-$2^n$  oscillators  whose trajectories ${\bf
c}_1(t),{\bf  c}_2(t)$  lie  on  the same attractor, but which
are  nevertheless  non-identical  since  at any given time $t$
their  dynamical  variables  are  different ${\bf c}_1(t) \neq
{\bf  c}_2(t)$.  Since the orbits are periodic there is a time
$\delta  t$  such  that  ${\bf  c}_1(t  +  \delta  t)  =  {\bf
c}_2(t)$   for   any  $t$.  This  operation  can  be  formally
considered   as  an  action  of  translation  operator  ${\cal
T}_{\delta t}$ on the trajectory of the first oscillator:
\begin{equation}
 {\cal T}_{\delta t}\; {\bf c}_1(t) = {\bf c}_1(t + \delta t) = {\bf c}_2(t).
\end{equation}
The  concentration  time  series  ${\bf  c}(t)$  of  the first
oscillator  then  appears to be shifted backward by $\delta t$
relative  to  that of the time series of the second oscillator
if   $\delta   t  >  0$  and  forward  otherwise.  Of  course,
trajectories  corresponding  to different attractors cannot be
made  to  correspond by such time translations, e.g. $P_{2^n}$
attractors  described  by different permutations $\pi^{(n)}_i$
have  different  patterns  of  oscillation,  but  even  if two
$P_{2^n}$  lie  in  the  same $\pi^{(n)}_i$ class their actual
shapes in ${\cal P}$ may differ significantly.

To  compare  the  local  dynamics  at  different points in the
medium   one   needs   to   single   out  the  most  important
characteristic  features  of  the  oscillation  pattern  while
discarding  unnecessary  details.  A  coarse-grained  symbolic
description  of  trajectories  appears  to  be useful for this
purpose.  We  assume  that  the times $t_1,t_2,\ldots,t_{2^n}$
at   which   the  trajectory  crosses  a  surface  of  section
$\phi=\phi_0$  (see  Sec.~2) are approximately equally spaced,
independent  of  the choice of $\phi_0$. Thus, the phase point
${\bf  c}(t)$  moving  along $P_{2^n}$ takes approximately the
same   time   $T_{2^n}/2^n$  to  traverse  each  loop  of  the
attractor.\cite{f5}  At  $t=t_0$  let  the  phase point of the
period-$2^n$  orbit  be  on the $j_0$-th loop of $P_{2^n}$, at
$t=t_0+T_{2^n}/2^n$  on  the  $j_1$-th  loop, and so on (where
$j_l\in[1,2^n],\;  l\in[1,2^n],\;  j\equiv\!\!\!\!\!\!/ \; l$)
until  at  $t=t_0+T_{2^n}$  the  phase  point  returns  to the
$j_0$-th  loop  and  the  pattern  $(j_0,j_1,\ldots  j_{2^n})$
repeats.  The  symbolic  string $s_j=(j_0,j_1,\ldots j_{2^n})$
constructed  in  this  way captures the most significant gross
features  of  the  oscillation  pattern  it describes. In this
coarse-grained   representation   the   number   of   possible
non-identical   trajectories  corresponding  to  a  particular
$\pi^{(n)}_i$   of  $P_{2^n}$  is  finite  and  the  different
trajectories   are   simply   given   by   the   $2^n$  cyclic
permutations   of   $s_j$.   Likewise   the  time  translation
operators   constitute   a  finite  group  ${\cal  T}_{l},  \;
l\in[-2^{n-1},2^{n-1})$.  They  act  on  the  symbolic  string
representing   the   orbit   to   give   one   of  its  cyclic
permutations. ~From  its definition it can be easily seen that
$\pi^{(n)}_i$     serves    as    a    symbolic    permutation
representation   of  ${\cal  T}_{+1}$  for  the  corresponding
$i$-th  permutation  class  of  $P_{2^n}$. Indeed, consider as
an  example  a period-4 oscillation whose representative point
lies  on  loop 3 at the reference moment of time $t=t_0$. Then
for  the  pattern  of  oscillation determined by $\pi^{(2)}_1$
the   state  reads  $s_1=(3241)$.  To  obtain  the  new  state
translated  by  $T_4/4$  backward  one  acts  on  $s_1$ by the
permutation   representation   $\pi^{(2)}_1$   of  the  ${\cal
T}_{+1}$ operator to get
\begin{equation}
	 {\cal T}_{+1}\: s_1=
	 \left(^{{\displaystyle 1234}}_{{\displaystyle 3421}}\right)
         (3241)=(2413)=s_2,
\end{equation}	 
which  correctly  describes  the  result  of  the shift of the
initial state $s_1$.

\section{ Global organization of medium}\label{glob}

\subsection{Period-1 regime}

We  now  return  to the spatially distributed medium and begin
by  reviewing  some  properties  of  the local dynamics in the
vicinity  of  a stable defect with topological charge $n_t=\pm
1$  in  a  period-1 oscillatory medium. Consider a cyclic path
$\Gamma=\{r=r_0>r_{max},\theta\in    [0,2\pi)\}$   surrounding
the  defect.  Here  $r_{max}$  is  a  radius such that for all
$(r,\theta),  r>r_{max},\:  \theta\in  [0,2\pi)$  the shape of
the  local  orbit  in phase space ${\cal P}$ is independent of
$(r,\theta)$  and  closely  approximates  that of the period-1
attractor    of    the    mass    action    rate    law   (see
Sec.~\ref{spiral}).  If  one  starts  at  an  arbitrary  point
$(r_0,\theta_0)\in  \Gamma$  one  finds that the instantaneous
local  phase  $\phi({\bf  r},t)$  changes by $2\pi$ or $-2\pi$
(depending  on  the  sign  of  the  topological  charge) along
$\Gamma$.  Let  us  now  fix a particular time instant $t=t^*$
and  construct  the  set  of  points  ${\cal S}=\{{\bf c}({\bf
r},t^*),r\in\Gamma\}$    as    a    phase   space   image   of
instantaneous concentrations at points lying on $\Gamma$.
\begin{figure}[htbp]
\begin{center}
\leavevmode
\epsffile{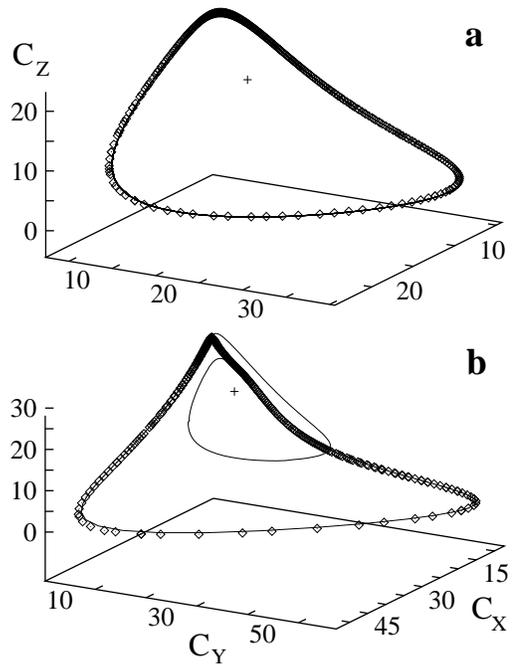}
\end{center}
\caption{${\cal  S}$-curves  (shown  by  diamonds) constructed
for   $\Gamma$   with   $r_0=55$   in   period-1   oscillatory
($\kappa_2=1.420$)  (a)  and  chaotic  ($\kappa_2=1.567$)  (b)
media.  Solid  curves  represent short time local trajectories
on $\Gamma$.}
\label{s-curve}
\end{figure}
The  property  of  a  defect (\ref{charge}) and the continuity
of  the  medium  insure  that  ${\cal  S}$  is a simple closed
curve      winding      once      around      ${\bf     c}^*$.
Figure~\ref{s-curve}(a)    shows    the    ${\cal    S}$-curve
constructed  for  the  contour $\Gamma$ with radius $r_0=55,\,
r_0>r_{max}$   in   a   period-1   oscillatory   medium   with
$\kappa_2=1.420$.   Since   all   the  points  on  the  ${\cal
S}$-curve  lie  at  the  same  time  on the local trajectories
$C({\bf   r}|t_0,\tau),\,   {\bf   r}\in\Gamma$  with  $t^*\in
[t_0,t_0+\tau]$,  and  for $\Gamma$ with $r_0>r_{max}$ all the
local  trajectories  are  the  same  and  approximated  by the
period-1  attractor  of  the  system  (\ref{mass}), the ${\cal
S}$-curve  simply  coincides with this attractor for any $t^*$
(cf.     Fig.~\ref{s-curve}(a)).    The    ${\cal    S}$-curve
constructed  for  an  arbitrary  simple closed path encircling
the  defect  in the medium possesses the same property as long
as the path lies in the open region $r>r_{max}$.

This   result   can   be   reformulated   in   terms  of  time
translations  of  local trajectories as follows. Let the local
trajectory    $C(r_0,\theta_0   |t_0,\tau)$   at   the   point
$(r_0,\theta_0)\in\Gamma$  be  taken  as a reference, then all
of   the  local  trajectories  on  $\Gamma$  can  be  obtained
through  the  translation  of  $C(r_0,\theta_0  |t_0,\tau)$ by
some  time  $\delta t(\theta-\theta_0)$ (see Sec.~\ref{topo}).
The    condition    (\ref{charge})    implies   that   $\delta
t(\theta-\theta_0)$     is    a    monotonically    increasing
(decreasing)  function  such  that  $\delta  t(2\pi)=\pm  T_1$
where  $T_1$  is  the  period  of  oscillation and the sign is
that  of  $n_t$. Thus, the oscillation pattern is continuously
time  shifted  along  $\Gamma$  such  that  upon return to the
initial point it has experienced translation by the period.

\subsection{Period-$2^n$ regime}

For  $2^n$-periodic  and chaotic media property (\ref{charge})
holds  where  $\phi({\bf  r},t)$  should  be understood as the
angle  variable  introduced  in Sec.~\ref{spiral}. This can be
seen  from  the  following  argument.  Take  a period-2 medium
with   rate   constants   chosen   in   the  vicinity  of  the
bifurcation   from   period-1   to   period-2  such  that  the
attractor  $P_2$  of  (\ref{mass})  lies infinitesimally close
to  $P_1$  from  which it bifurcated. Due to the continuity of
the    solutions    of    the    reaction-diffusion   equation
(\ref{rds}),  the  value  of $\oint \phi({\bf r},t)\;d{\bf l}$
cannot  change  abruptly  when  the  bifurcation  parameter is
changed   through   the   period-doubling   bifurcation.  This
implies  that  the  ${\cal S}$-curve constructed for a contour
$\Gamma$  in  a  period-$2^n$  medium,  as in case of a simple
period-1  medium,  is  a  closed curve which loops once around
${\bf  c}^*$  in phase space. This is illustrated in panel (b)
of  Fig.~\ref{s-curve}  which  shows  ${\cal  S}$  for contour
$\Gamma$    with    radius    $r_0=55$    in    medium    with
$\kappa_2=1.567$  and  time  $t=t^*$.  Recall  again  that the
points  of  the  ${\cal  S}$  curve  have  to lie on the local
trajectories  $C({\bf  r}|t_0,\tau),\,  {\bf r}\in\Gamma$ (cf.
Fig.~\ref{circ}  where  points  designated  by diamonds lie on
${\cal  S}$  for  the  chosen time moment and contour shown in
the  figure).  Since  the local trajectories in a period-$2^n$
medium  loop  several  times  around  ${\bf  c}^*$,  the curve
${\cal  S}$  which  winds  only once $(n_t=\pm1)$ around ${\bf
c}^*$  cannot  span the entire local trajectory as is the case
for     a     period-1    medium.    As    one    sees    from
Fig.~\ref{s-curve}(b)  ${\cal  S}$  follows the larger loop of
the  local  trajectory,  which  for  $\Gamma$ with $r_0=55$ is
typically   a   period-2   orbit  (cf.  Fig.~\ref{circ}),  and
instead  of  making  the  second  turn on the smaller loop, it
crosses  the  gap  between  the  loops  and  closes on itself.
Although  the  shape  of  ${\cal  S}$  changes  with time (see
\cite{prl}  for  details),  for any $t^*$ there exist segments
of   ${\cal   S}$  which  connect  different  loops  of  local
trajectories.  This  behavior  of  the ${\cal S}$ curves would
be   impossible   if  loop  exchanges  were  nonexistent.  The
analysis  shows  that  the segments of ${\cal S}$ covering the
gaps  between  the  loops of the local trajectories are images
of  points  on  $\Gamma$  which  lie close to the intersection
with  the  $\Omega$  curve.  Thus, the loop exchanges observed
in   period-$2^n$   media   are  necessary  to  reconcile  the
contradiction  between  the  one-loop  topology  of the ${\cal
S}$  curves  determined  by  the  presence of a defect and the
multi-loop  topology  of  the local trajectories determined by
the local rate law.

The  change  of  the  local  trajectories  along  the  contour
$\Gamma$  in  period-$2^n$ media can be considered in terms of
time  translations  if  one  adopts  a  generalization  of the
translation  operation  in  the  following  way. In a period-2
medium  let  the  contour  $\Gamma$  and  the  reference point
$(r_0,\theta_0)\in\Gamma$   be   chosen   so   that   $\Gamma$
intersects   the   $\Omega$   curve   in   the   single  point
$(r_0,\theta_{\Omega})$  and  suppose  that  these  points are
sufficiently  separated  from  each other. Since the shapes of
the  local  orbits  change significantly along any closed path
surrounding    a    defect    (cf.    Fig.~\ref{circ})   these
trajectories  cannot  be  made to coincide by time translation
as    this    operation   is   defined   in   Sec.~\ref{topo}.
Nevertheless,  the  general  features  of the temporal pattern
of  the  trajectories  are  preserved  (e.g.  sharp  maxima in
$c_i(t)$ time series) and for two
\begin{figure}[htbp]
\begin{center}
\leavevmode
\epsffile{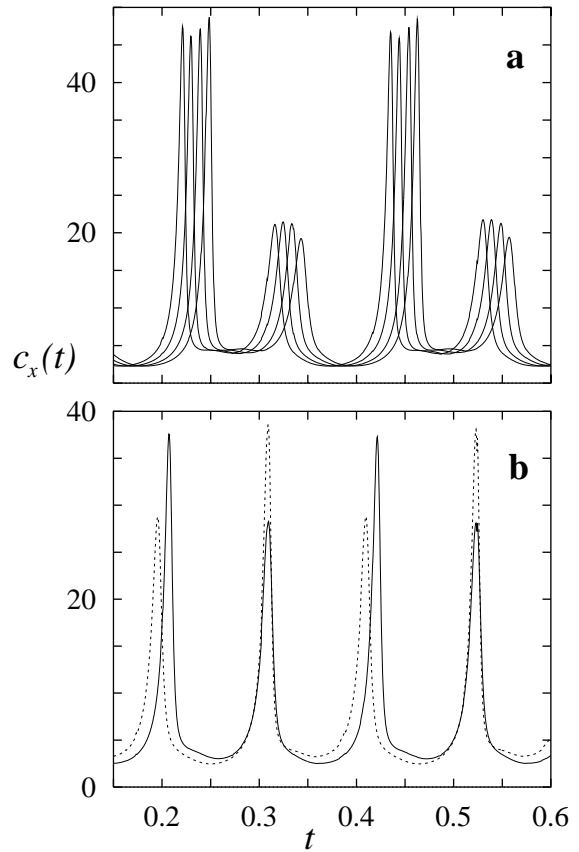}
\end{center}
\caption{Period-2  local  concentration  time series $c_x({\bf
r},t)$  calculated  on  a cyclic path $\Gamma$ surrounding the
defect:  (a)  series  sampled  at  four  consecutive locations
separated  by  $\delta\theta=30^o$;  (b) two series sampled at
locations   chosen   symmetrically   on  either  side  of  the
intersection with the $\Omega$ curve. }
\label{gamma}
\end{figure}

\noindent   locations  $(r_0,\theta_1)$  and  $(r_0,\theta_2)$
one    is    able    to    find    a    time   shift   $\Delta
t(\theta_1,\theta_2)$   such   that   some   measure   of  the
deviation between the trajectories, say,
\begin{equation}
\label{match}
M(\Delta t(\theta_1,\theta_2))= \int_{t_0}^{t_0+\tau} 
|{\bf c}^{(1)}(t+\Delta t) - {\bf c}^{(2)}(t)|\: dt ,
\end{equation}
is     minimized.     Choosing     the     local    trajectory
$C(r_0,\theta_0|t_0,\tau)$  as  a  reference  and comparing it
to  all  the  other  local  orbits  on $\Gamma$ one is able to
define      the      time      shift      function     $\delta
t(\theta-\theta_0)\equiv\Delta     t(\theta,\theta_0)$.    The
shift   function  $\delta  t(\theta-\theta_0)$  increases  (or
decreases)    monotonically    and    almost   linearly   (see
Fig.~\ref{gamma}(a))  with  $d  (\delta  t)/d  \theta  \approx
T_2/2\cdot  2\pi$  everywhere  on  $\Gamma$ except for a small
neighborhood  of  $\theta=\theta_{\Omega}$  where  it exhibits
break.  Indeed,  the loop exchange at $\theta=\theta_{\Omega}$
causes  the  discontinuity  of $\delta t(\theta-\theta_0)$. At
$\theta=\theta_{\Omega}$   both   loops  of  the  local  orbit
become   equivalent   and   the   oscillation  is  effectively
period-1  with  period  $T_1=T_2/2$.  Since the loops exchange
at   $\theta=\theta_{\Omega}$,   to   find   the   best  match
(\ref{match})  between  local  trajectories  sampled at points
$(r_0,\theta_{\Omega}            -\varepsilon)$            and
$(r_0,\theta_{\Omega}+\varepsilon)$,  one  needs  to translate
one   of   the   trajectories   by   $\delta   t   =   T_1   +
O(\varepsilon)$.     This    can    be    easily    seen    in
Fig.~\ref{gamma}(b)   which   displays   two  $c_x(t)$  series
calculated        at        spatial        points        lying
$\theta-\theta_{\Omega}=\pm    10^o$   on   either   side   of
$\theta_{\Omega}$ on $\Gamma$.

\subsection{ Trajectory transformations along $\Gamma$}

The  transformation  of  local trajectories along $\Gamma$ can
be  imagined  to  occur  a  result  of two separate processes.
Suppose        everywhere       on       $\Gamma$       except
$\theta=\theta_{\Omega}$  the  shape of the local trajectories
in  ${\cal  P}$  is  the  same  and  is  equivalent to that of
$C(r_0,\theta_0|t_0,\tau)$.   Then   all   the   other   local
trajectories  $C(r_0,\theta|t_0,\tau),  \;  \theta\in[0,2\pi),
\theta\neq  \theta_{\Omega}$  can be found by time translation
of         $C(r_0,\theta_0|t_0,\tau)$        by        $\delta
t(\theta-\theta_0)=T_2(\theta-\theta_0)/2\cdot  2\pi$.  Assume
that  all  the deformations of the phase space portrait of the
local  trajectory  which  take place along $\Gamma$, including
the    exchange    of    loops,    occur    at    the    point
$\theta=\theta_{\Omega}$   so   that   the   passage   through
$\theta_{\Omega}$    shifts   the   oscillation   by   $\delta
t_p=T_1=T_2/2$.   Then  the  result  of  the  continuous  time
translation   that  occurs  during  $2\pi$  circulation  along
$\Gamma$  may  be  described  by  the  action  of  the  ${\cal
T}_{n_t}$  operator  $(n_t=\pm  1)$,  while  the result of the
loop   exchange   is   described   by   the   operator  ${\cal
T}_{-n_t}$.~\cite{f9}  The  total  transformation of the local
oscillation   after   a   complete   cycle  over  $\Gamma$  is
equivalent   to  the  identity  transformation  and  thus  the
result  is  in  accord  with  continuity of the medium. If one
makes  the  assumption  that  loop  exchange does not occur on
some  contour  $\Gamma$  encircling  a  defect with $|n_t|=1$,
the  time  shift  function $\delta t(\theta-\theta_0)$ becomes
monotonic  and  continuous everywhere on $\Gamma$. As a result
one  arrives  at  the  incorrect conclusion that starting from
the   point  $(r_0,\theta_0)$  with  the  oscillation  pattern
symbolically    represented   by   the   string   $s_1$,   say
$s_1=(12)$,   and  moving  along  $\Gamma$  in  the  clockwise
direction  one  returns  to  the  same  point  $(r_0,\theta_0+
2\pi)\equiv  (r_0,\theta_0)$  but with the oscillation pattern
shifted  by  $T_2/2$  and  given  by  $s_2=(21)\neq s_1$. Note
that  this  contradiction  does  not  arise  in  the  period-1
oscillatory  medium  where  circulation  over  any closed path
encircling  a  defect results in the translation by the entire
period    which   automatically   satisfies   the   continuity
principle.   Thus   the   necessity   of   loop  exchanges  in
period-$2^n$,   $n>0$   media   with   a   topological  defect
demonstrated  earlier  in  this  section  in  terms  of ${\cal
S}$-curves is now explained in terms of time translations.

The  results  for  the  period-2 medium can be generalized for
any  $n>1$  using  the  following  hypothesis. ~From  the main
property  of  a  topological  defect (\ref{charge}) it follows
that   integration   of   an  infinitesimal  continuous  shift
$d(\delta  t)$  over  any  closed  path  surrounding  a defect
results  in  a  total  shift  by  $\pm T_{2^n}/2^n$ and can be
symbolically  described  by  the  ${\cal  T}_{n_t}$  operator.
Numerical   simulations  demonstrate  the  existence  of  time
translation  discontinuity  points such that sum of $\delta t$
jumps   over   these   points  amounts  to  a  shift  of  $\mp
T_{2^n}/2^n$  described  by  the  ${\cal  T}_{-n_t}$ operator.
The   locations   of   these  points  in  the  medium  can  be
identified  with  the  $\Omega$  curve  and  the origin of the
time   translation  discontinuities  with  the  loop  exchange
phenomenon.   The   relation   (\ref{prod})  of  the  Appendix
connects  translations  and  loop  exchanges and allows one to
predict  the  number  and the kind of loop exchanges necessary
to perform the required ${\cal T}_{-n_t}$ translation.
	    
\subsection{Examples}

\begin{figure}[htbp]
\begin{center}
\leavevmode
\epsffile{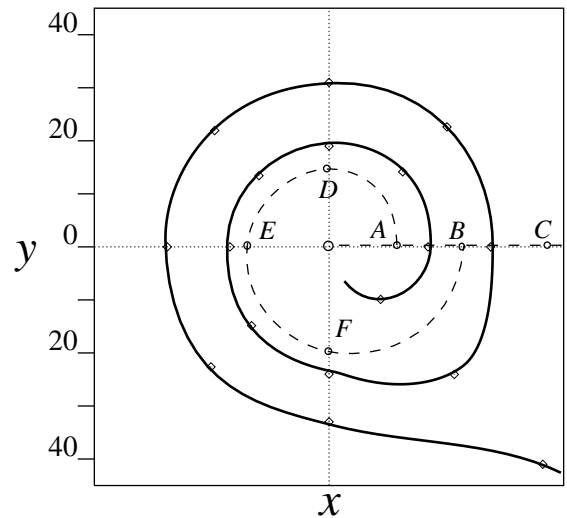}
\end{center}
\caption{Sketch  of  the  $\Omega$  curve for the square array
($\kappa_2=1.567,\;L=80$).   The  points  were  obtained  from
simulations.  The  ray  $ABC$ intersects $\Omega$ at locations
with radii 20 and 31.}
\label{s-om}
\end{figure}

Consider   again  the  square  $80\times80$  array  with  rate
constants       corresponding      to      chaotic      regime
($\kappa_2=1.567$).   As   period-4   fine  structure  is  the
highest  level  of  local organization resolved in the medium,
it  is  sufficient  to  use  the formalism developed above for
$P_4$  to  describe  the  local  dynamics.  The analysis shows
that  in  the  bulk  of the medium the oscillation is given by
the  $\pi^{(2)}_1=(^{1234}_{3421})$  pattern.~\cite{f10} Using
this  data  and  the results presented in the Appendix one can
easily  enumerate  all the sequences of exchanges resulting in
${\cal   T}_{+1}$   translation.  Indeed,  one  should  expect
either  exchange  of loops 3 and 4 followed by the exchange of
period-2   bands  $(^{1234}_{3412})$  or  first  the  period-2
bands  exchange  followed  by  exchange  of  loops  1  and  2.
Figure~\ref{s-om}   is   a  schematic  representation  of  the
medium  with  a  negatively  charged  $(n_t=-1)$ defect in the
center and the $\Omega$ curve displayed.

Consider  the  change  of  the  oscillation  pattern along ray
$ABC$   emanating   from  the  defect  as  the  value  of  $r$
increases    (see    Fig.~\ref{s-frm}   for   the   cumulative
first-return  map  constructed  for  this ray). The pattern of
oscillation    $s_A=(4132)$   corresponding   to   permutation
$\pi^{(2)}_1=(^{1234}_{3421})$  can  be followed from $r=5$ to
$r=19$   where  the  period-2  bands  undergo  exchange.  This
results  in  the  switch  to the oscillation pattern described
by   $\pi^{(2)}_2=(^{1234}_{4312})$   seen   at   $r=21$.  The
pattern   $\pi^{(2)}_1$  is  restored  after  loops  1  and  2
exchange  at  $r=22$  and  and  this  pattern  persists  until
another exchange occurs at $r=28$.

\begin{figure}[htbp]
\begin{center}
\leavevmode
\epsffile{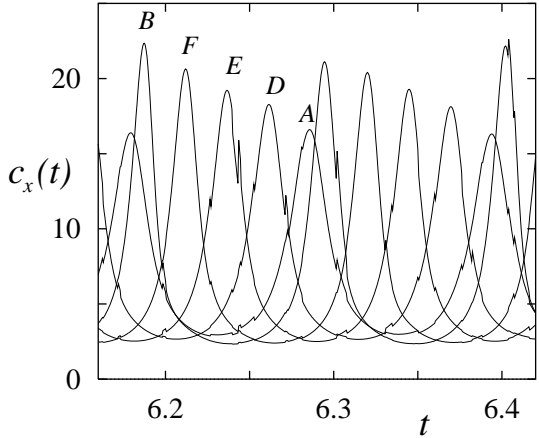}
\end{center}
\caption{Concentration    time    series    $c_x({\bf   r},t)$
calculated  at  points  $A,D,E,F,B$  of the square array shown
in Fig.~15. }
\label{abc}
\end{figure}

Using  translation  operator  ${\cal  T}_{+1}$ one can express
the  transition  of  the  state  $s_A$  ($r<20$)  through  the
sequence  of  loop  exchanges  described  above  to  the state
$s_B=(1324)$  (for  $22<r<28$) as $s_B={\cal T}_{+1} s_A$. The
same  shift  can  be  achieved by continuous translation along
the  path  $ADEFB$ which does not intersect $\Omega$ but winds
once  counter-clockwise  around  the defect. The $c_x(t)$ time
series   at   points   $A,D,E,F$  and  $B$  are  displayed  in
Fig.~\ref{abc} and demonstrate that this is the case.

Continuing  to  advance along the ray $ABC$, one finds that at
$r=28$  loops  3  and 4 exchange and oscillation switches once
more  to  the  state corresponding to $\pi^{(2)}_2$. After the
period-2  band  exchange  at  $r=32$ the pattern corresponding
to  $\pi^{(2)}_1$  is reinstated and remains unchanged for all
$r>32$.   Again   the   oscillation   at   $r>32$,   described
symbolically   by  $s_C=(3241)$,  appears  to  be  shifted  by
$T_4/4$ relative to $s_B$ and by $T_4/2$ relative to $s_A$.

\begin{figure}[htbp]
\begin{center}
\leavevmode
\epsffile{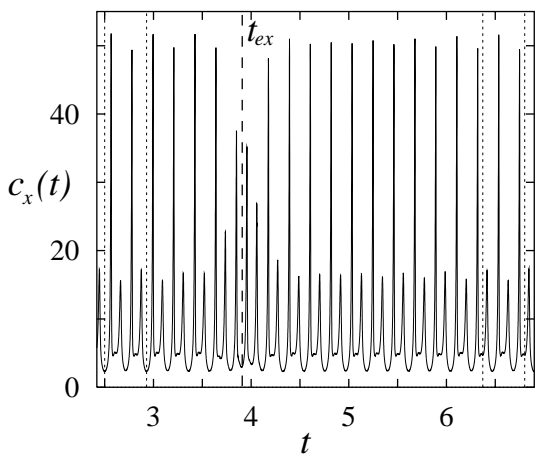}
\end{center}
\caption{   Segment   of   the   concentration   time   series
$c_x(r,t)$    calculated    for    the    disk-shaped    array
($\kappa_2=1.567,\;R=80$)  at  $r=76$ showing the $T_4/4$ time
shift  of  the  oscillation  pattern  (see  explanation in the
text).}
\label{p4ex}
\end{figure}

The  existence  of a $T_4/4$ shift after crossing $\Omega$ can
also  be  seen from the results for the disk-shaped array with
$R=80$.  Figure~\ref{p4ex}  shows  a  segment of the $c_x({\bf
r},t)$  time  series  sampled in a fixed frame $(r,\theta)$ at
$r=76$.  In  this  coordinate  system  $\Omega$ slowly rotates
clockwise  (again  $n_t=-1$)  with  period  $T_{ex}$. Two time
windows  each  of  length  $T_4$  marked  by  dotted lines and
separated  by  $\Delta  t  =  8T_4$  allow  one to see how the
oscillation   state  (4132)  is  substituted  by  its  forward
$T_4/4$  translation  (2413)  after  the $\Omega$ curve passes
the observation point at $t = t_{ex}$.

\section{Conclusions} \label{conc}      		    

General    principles    underlie    the    organization    of
$2^n$-periodic  or  chaotic  media supporting spiral waves. As
in  simple  oscillatory  media,  the  core  of  a  spiral is a
topological   defect   which  acts  as  an  organizing  center
determining   dynamics   in   its   vicinity;   however,   the
structural  organization  of  the  medium that arises from the
existence  of  the  defect is far more complicated. Due to the
absence   of   a   conventional   definition   of   phase  for
oscillations  more  complex  than period-1, the identification
of  a  defect  in  terms of the relation (\ref{charge}) is not
obvious    and    requires    the   introduction   of   (often
model-dependent)  phase  substitutes  which  for  some systems
may    be   provided   by   angle   variables.   Despite   the
complications  with  the definition of phase, one can identify
a  defect  in  terms  of  local  trajectories.  Indeed, as one
moves  away  from the defect the local dynamics takes the form
of   a   progression   of  period-doubled  orbits,  from  near
harmonic,   small-amplitude,   period-1  orbits  to  ``noisy''
period-$2^l$  orbits,  where  $l$  is  a function of variables
such as diffusive coupling and the system size and shape.

The  presence  of  a defect imposes topological constraints on
the  global  organization  of  medium  as  well.  As was shown
above,  when  $2^n>n_t$  the $2^n$-loop structure of the local
trajectories  conflicts  with  the  period-$n_t$  structure of
the  ${\cal  S}$-curve  and  a  complex,  asymmetric,  spatial
pattern   of   local  dynamics,  the  defect-organized  field,
arises   as   a  result  of  the  necessity  to  maintain  the
continuity  of  the  medium. The most prominent characteristic
feature  of  this  field  is the $\Omega$ curve defined as the
set   of   points   where  the  local  dynamics  most  closely
resembles  period-1.  This  signals  the  exchange of period-2
bands.  If  the  local  trajectories  possess  structure finer
than  period-2,  other  loop  exchanges leading to more subtle
changes  in  the  local orbits can be found in the vicinity of
$\Omega$.  The  net  result of these exchanges is to produce a
time  shift  of  the  trajectories  which  compensates for the
smooth  time  translation  accumulated  on  continuous  paths.
Since  the  topological  continuity  must  be  observed on any
arbitrarily   large  closed  path  encircling  a  defect  and,
therefore,  this  contour  has  a  point  of intersection with
$\Omega$,  a  single  defect  in a period-$2^n$ $(n>0)$ medium
cannot be localized.

We  point  out  again  that  many  of  the  phenomena  we have
discussed  above  are  not  dependent  on  the  existence of a
period-doubling  cascade  or  chaotic local dynamics, although
this    is    the   case   we   have   analysed   in   detail.
Reaction-diffusion   systems   with   local  complex  periodic
orbits  in  phase  space  dimensions  higher  than  two should
exhibit  similar  features  when they support spiral waves. It
should  be  possible  to  experimentally  probe  the phenomena
described  in  this  paper.  The  appropriate parameter regime
can   be   determined   from  investigations  of  well-stirred
systems.  For  example, period-doubling and chaotic attractors
have  been  observed  in  the  Belousov-Zhabotinsky  reaction.
\cite{BZchaos}  If  the  spiral  wave dynamics is then studied
in   a  continuously-fed-unstirred  reactor  \cite{cfur},  one
should  be  able  to observe the characteristics of the spiral
dynamics   and   the  loop  exchange  process  that  serve  as
signatures of the phenomena described above.

\section{Acknowlegements}       
We  thank  Peter  Strizhak  for  his interest in this work and
helpful  comments.  This work was supported in part by a grant
from  the  Natural  Sciences  and Engineering Research Council
of Canada and by a Killam Research Fellowship (R.K.).

\begin{appendix}
\section{Braid moves and loop exchange operators}
       
In  this  appendix  we  make  use  of  the  projection  of the
period-doubled   attractors   $P_{2^n}$   onto  closed  braids
${\bar  B}_{2^n}$  (see  Sec.~\ref{topo})  to  show  how  loop
exchanges  affect  the  pattern of oscillation. We demonstrate
that   those  combinations  of  loop  exchanges  that  produce
identity  transformations  of  $P_{2^n}$  result in nontrivial
time translations of trajectories.

\begin{figure}[htbp]
\begin{center}
\leavevmode
\epsffile{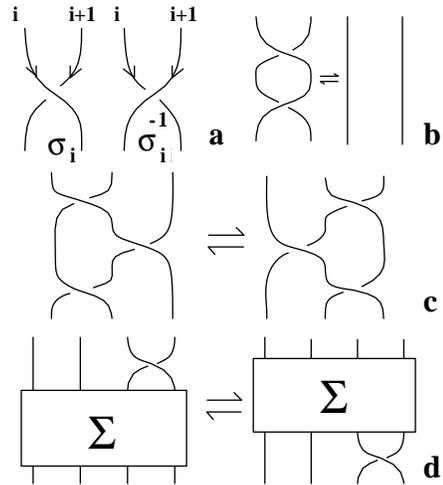}
\end{center}
\caption{Conventional  designations  and  basic  braid  moves:
(a)   definition   of   elementary   braids   $\sigma_i$   and
$\sigma_i^{-1}$;  (b)  type  2  Reidemeister  move; (c) type 3
Reidemeister   move,  and  (d)  first  Markov  move  ($\Sigma$
represents arbitrary braid). }
\label{brth}
\end{figure}

Each  closed  braid  ${\bar  B}_{2^n}$ is represented by a set
of   non-identical  braid  words  with  their  number  rapidly
growing  with  $n$.  Without  violation  of  the  topology  of
${\bar  B}_{2^n}$  they can be transformed into one another by
the following set of moves (see, e.g. \cite{bir}) :
\begin{enumerate}
\item     commutation     relation,     $\sigma_i\sigma_j    =
\sigma_j\sigma_i, \;\; |i-j| \geq 2$;
\item  type  2  Reidemeister  move,  $\sigma_i\sigma_i^{-1}  =
\sigma_i^{-1}\sigma_i = {\bf 1}$;
\item         type         3         Reidemeister        move,
$\sigma_i\sigma_{i+1}\sigma_i                                =
\sigma_{i+1}\sigma_i\sigma_{i+1}$;
\item   first   Markov  move,  $\sigma_i\Sigma\sigma_i^{-1}  =
\sigma_i^{-1}\Sigma\sigma_i = \Sigma, \; \Sigma \in B$;
\end{enumerate}
where  $B$  is  a  set  of  open braids. While the first three
rules  are  common  for  all  braids,  rule  4 is specific for
closed   braids.  Indeed,  it  can  be  written  in  the  form
$\sigma_i\Sigma   =   \Sigma\sigma_i$  which,  for  elementary
braids  $\sigma_i$,  corresponds  to  moving $\sigma_i$ around
${\bar  B}_{2^n}$  resulting  in  the  exchange  of the closed
braid  loops  (cf.  Fig.~\ref{brth}(d)).  Type  1 Reidemeister
(or  second  Markov)  moves  are not allowed since they do not
preserve   the  number  of  loops,  an  essential  feature  of
$P_{2^n}$

\noindent  attractors.  While  rules  1, 2 and 3 do not affect
$\pi^{(n)}_i$,  the  first  Markov  move  does (except for the
degenerate  case  of  $P_2$ which is represented by the single
permutation     $\pi^{(1)}_1$).    Thus    any    number    of
rearrangements  affecting  only  the  braid  $B_{2^n}$  of the
closed  braid  ${\bar  B}_{2^n}$  leave  the braid word in the
same  permutation  class $\pi^{(n)}_i$, while each application
of the first Markov move yields a new permutation class.
			    
\subsection{Loop exchanges for $P_2$ and $P_4$}

We  now  examine how the loop exchanges influence the patterns
of  oscillation  for the period-2 and period-4 attractors. For
$P_2$   one   has   only   the  single  braid  word  $\sigma_1
(\sigma_1^{-1})$  and  the  single  permutation $\pi^{(1)}_1 =
(^{12}_{21})$  induced  by  $\sigma_1$. Two different symbolic
states  $s_1  =  (12)$  and  $s_2 = (21)$ are possible for the
period-2  oscillation  with  respect to some fixed time frame.
We  introduce  an  operator  $A^{(1)}_1$  whose  action on the
closed  braid  representing  $P_2$  is  to  move $\sigma_1$ by
$2\pi$  in  a  direction  opposite  to the flow. The result of
the  action  of  this operator, which is the first Markov move
for  $\Sigma  =  {\bf  1}$,  is  to  leave the attractor $P_2$
unchanged;  however,  one  finds  that  loops  1  and  2  have
exchanged  their  locations in phase space. In the time series
for  the  dynamical variable $c_i(t)$ the exchange can be seen
as  a  substitution of taller maxima (2) by shorter maxima (1)
and  vice-versa.  If  this  process  is  followed  in  time it
produces     the     characteristic     pattern    shown    in
Fig.~\ref{p2ex}.  Thus,  application  of  $A^{(1)}_1$ to $P_2$
induces  a  transformation of the oscillation state $s_1$ into
$s_2$  and  vice-versa.  This can be symbolically described as
an   action   of   an  exchange  operator  ${\cal  A}^{(1)}_1$
represented by the permutation $(^{12}_{21})$ :
\begin{eqnarray}
 {\cal A}^{(1)}_1\: s_1 & = & 
 \left(^{{\displaystyle 12}}_{{\displaystyle 21}}\right)
 (12) = (21) = s_2, \nonumber \\
 {\cal A}^{(1)}_1\: s_2 & = & 
 \left(^{{\displaystyle 12}}_{{\displaystyle 21}}\right)(21) = (12) = s_1. 
\end{eqnarray} 
One  sees  that the action of ${\cal A}^{(1)}_1$ is equivalent
to  that  of  ${\cal T}_{+1}$ which translates the oscillation
pattern  by  half  a period. The inverse of the braid operator
$A^{(1)}_1$  can  be  introduced  in  an  analogous  way as an
operator  moving  $\sigma_1$  {\it along} the direction of the
flow.   It   corresponds   to  an  exchange  operator  $({\cal
A}^{(1)}_1)^{-1}  =  {\cal  T}_{-1}$  acting on strings. Since
for  $P_2$  application  of ${\cal T}_{+1}$ or ${\cal T}_{-1}$
results  in  essentially the same states the sign of the shift
is   chosen   to   maintain   consistency  with  corresponding
operators  for  $P_{2^n}$, $n>1$. Double application of ${\cal
A}^{(1)}_1$  results  in translation by a full period and thus
in the identity operator
\begin{equation}
 ({\cal A}^{(1)}_1)^2 = ({\cal A}^{(1)}_1)^{-2} = {\bf 1}.
\end{equation} 

The    $P_4$    attractor    possesses   a   richer   set   of
transformations.   Since  under  coarse-graining  braid  words
which  induce  the  same  $\pi^{(n)}_i$ are indistinguishable,
we  can  single  out two essential representatives $\Sigma_1 =
\sigma_3\Sigma$    for    $\pi^{(2)}_1=(^{1234}_{3421})$   and
$\Sigma_2      =     \sigma_1\Sigma$     for     $\pi^{(2)}_2=
(^{1234}_{4312})$,      where      $\Sigma$     stands     for
$\sigma_2\sigma_1\sigma_3\sigma_2$.   Let  $A^{(2)}_1$  be  an
operator   on  ${\bar  B_2}$  which  moves  the  double-thread
crossing   (cf.   large   dashed   box  in  Fig.~\ref{p-4toB})
$\Sigma$  by  $2\pi$  in  the  direction opposite to the flow,
analogous  to  the  action  of  $A^{(1)}_1$  on $\sigma_1$ for
${\bar  B_2}$.  In fact, the action of $A^{(2)}_1$ can be seen
as   an   exchange  of  the  period-2  bands  of  $P_4$,  each
consisting  of  two  period-4  loops.  Unlike  $A^{(1)}_1$ the
operator   $A^{(2)}_1$   alternates   braid   words   and  the
corresponding           pattern-defining          permutations
$\Sigma_1,\pi^{(2)}_1
\stackrel{A^{(2)}_1}{\longleftrightarrow}
\Sigma_2,\pi^{(2)}_2$.    The   application   of   $A^{(2)}_1$
induces   transformations  of  symbolic  strings  $s_j$  which
again  can  be described by the action of an exchange operator
${\cal    A}^{(2)}_1$    represented    by   the   permutation
$(^{1234}_{3412})$.  Due  to the apparent similarity of action
of   $A^{(1)}_1$   and   $A^{(2)}_1$   on   the  corresponding
attractors,   ${\cal   A}^{(2)}_1$   inherits   the  algebraic
properties  of  ${\cal A}^{(1)}_1$. Indeed, ${\cal A}^{(2)}_1$
produces  identity  operator  when applied twice and, thus, is
equal to its inverse.

Finer  rearrangement  of  the $P_4$ loop structure is provided
by  the  action  of  $A^{(2)}_2$  defined as an operator which
moves  the  single  crossing  (enclosed  in the smaller box in
Fig.~\ref{p-4toB})  by  $2\pi$  in  the  direction opposite to
flow. ~From  the structure of ${\bar B_4}$ one sees that after
application  of  $A^{(2)}_2$  the  single  crossing  does  not
return  to  the same location in ${\cal P}$ but appears on the
other  period-2  band;  thus,  braid  words  and $\pi^{(n)}_i$
permutations  alternate.  Depending  on the initial state, the
application   of   $A^{(2)}_2$   results   in  different  loop
exchanges.  In  the case of $\Sigma_1 = \sigma_3\Sigma$ action
of  $A^{(2)}_2$  leads  to  the  exchange of loops $3$ and $4$
and  results  in  the  string  transformation described by the
exchange    operator    ${\cal   A}^{(2)}_2$   with   symbolic
representation  $(^{1234}_{1243})$.  When it acts on $\Sigma_2
=   \sigma_1\Sigma$  it  exchanges  loops  1  and  2  and  the
permutation  representation  of  ${\cal A}^{(2)}_2$ changes to
$(^{1234}_{2134})$.  The  inverse  of  $A^{(2)}_2$  moves  the
single   crossing   along   the  flow  and  produces  opposite
results;   i.e.,   acting   on   $\Sigma_1$  it  leads  to  $1
\leftrightarrow  2$  exchange  and  when applied to $\Sigma_2$
it results in the exchange $3 \leftrightarrow 4$.

Note  the  difference  between action of $A^{(n)}_i$ on braids
and  the  action  of  exchange operators ${\cal A}^{(n)}_i$ on
symbolic   strings.   While   several  operations  $A^{(n)}_i$
applied  to  the  same  initial  braid word lead to equivalent
final  words  (e.g. $A^{(2)}_1$ and $A^{(2)}_2$) the resulting
loop  exchanges  and, thus their permutation descriptions, can
be      quite      different      ($(^{1234}_{3412})$      and
$(^{1234}_{1243})$  for  the  example  chosen).  Consequently,
compositions  of  braid  operations returning the braid ${\bar
B}_{2^n}$   to   its   initial   state  (e.g.  $A^{(2)}_2\circ
A^{(2)}_2$)  may  induce  nontrivial translations of $s_j$. To
demonstrate  this  let $({\cal A}^{(2)}_2)^2$ act on the trial
state $s_1 = (3241)$. Applying the rules one obtains
\begin{equation}
 ({\cal A}^{(2)}_2)^2 \: s_1=
     \left(^{{\displaystyle 1234}}_{{\displaystyle 2134}}\right)
     \left(^{{\displaystyle 1234}}_{{\displaystyle 1243}}\right)
     (3241)=(4132)={\cal T}_{+2}\:s_1,
\end{equation}     
thus     relating     the     simultaneous    loop    exchange
$(13)\leftrightarrow  (24)$  with  translation of the period-4
oscillation  by  half  of  a period. This implies as well that
application  of  ${\cal  A}^{(2)}_2$ four times results in the
identity  string  transformation  $ ({\cal A}^{(2)}_2)^4={\cal
T}_{+4}={\bf       1}$       and,      therefore,      $({\cal
A}^{(2)}_2)^{-1}=({\cal    A}^{(2)}_2)^3$.   Compositions   of
braid  operators  $A^{(2)}_1$  and $A^{(2)}_2$ provide another
example  of  how  identity  braid  operators induce nontrivial
string   transformations.   Since  both  operators  and  their
inverses   alternate  braid  words  $\Sigma_1  \leftrightarrow
\Sigma_2$  the  application  of  the composition of any two of
them  returns  the braid to the same $\pi^{(2)}_i$ permutation
class  and,  thus,  the  resulting  string  transformation  is
equivalent   to   some  translation.  The  relations  for  the
compositions    of   the   ${\cal   A}^{(2)}_1$   and   ${\cal
A}^{(2)}_2$  operators  can  be  obtained  directly from their
symbolic representations :
\begin{eqnarray}		   
   \label{exch}
   {\cal A}^{(2)}_1 &\circ& {\cal A}^{(2)}_2 = 
   \left(^{{\displaystyle 1234}}_{{\displaystyle 1243}}\right)
   \left(^{{\displaystyle 1234}}_{{\displaystyle 3412}}\right) = 
   {\cal A}^{(2)}_2\circ {\cal A}^{(2)}_1 
   \nonumber \\ & &
   = 
   \left(^{{\displaystyle 1234}}_{{\displaystyle 3412}}\right)
   \left(^{{\displaystyle 1234}}_{{\displaystyle 2134}}\right) =
   \left(^{{\displaystyle 1234}}_{{\displaystyle 3421}}\right)
   = \pi^{(2)}_1 = {\cal T}_{+1} \; ,
             \\ 
   {\cal A}^{(2)}_1 &\circ& ({\cal A}^{(2)}_2)^{-1} = 
   \left(^{{\displaystyle 1234}}_{{\displaystyle 2134}}\right)
   \left(^{{\displaystyle 1234}}_{{\displaystyle 3412}}\right)=
   ({\cal A}^{(2)}_2)^{-1}\circ {\cal A}^{(2)}_1 
   \nonumber \\ & &
   = \left(^{{\displaystyle 1234}}_{{\displaystyle 3412}}\right)
     \left(^{{\displaystyle 1234}}_{{\displaystyle 1243}}\right)=
     \left(^{{\displaystyle 1234}}_{{\displaystyle 4312}}\right)=
     \pi^{(2)}_2 = {\cal T}_{-1} \; .
   \nonumber 
\end{eqnarray}
These  relations  are  constructed  using  the assumption that
the  initial  state  of  the  braid  is  $\Sigma_1$.  Although
application   to  an  alternative  initial  condition  changes
actual  permutation  representations of the exchange operators
it    yields    algebraically    equivalent    results.  ~From
(\ref{exch})   one   sees  that  all  the  exchange  operators
commute   and   their  compositions  provide  operators  which
translate  the  oscillation  by  all  the allowed multiples of
$T_4/4$.

\subsection{Loop exchanges for $P_{2^n}$ attractors}
  	      
\begin{figure}[htbp]
\begin{center}
\leavevmode
\epsffile{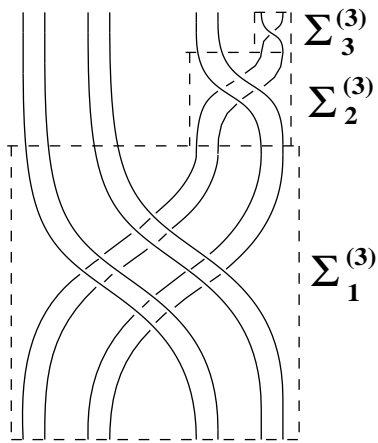}
\end{center}
\caption{Braid $B_8$ constructed for the $P_8$ attractor. }
\label{p-8}
\end{figure}

A   generalization   of   the  phenomena  discussed  above  to
arbitrary  $n$  may  be  inferred  from the observation of the
structural   organization   of   the   closed   braids  ${\bar
B_{2^n}}$    corresponding    to   period-doubled   attractors
$P_{2^n}$.  Indeed,  ${\bar B_{2^{n+1}}}$ can be obtained from
${\bar  B_{2^n}}$  by doubling each thread of ${\bar B_{2^n}}$
and  adding  a  single  crossing  on  top  to  preserve simple
connectivity  of  the construction. The braid ${\bar B_{2^n}}$
arising  as  a  result  of  $n$  successive iterations of this
procedure   can   be   subdivided   into  $n$  non-overlapping
structurally  similar  blocks  of  braids  $\Sigma^{(n)}_m, \;
m=\overline{1,n}$.  This  principle of structural organization
is  illustrated  in  Fig.~\ref{p-8} representing $B_8$ and its
three  crossing  blocks  shown  in  a  series  of  boxes  with
decreasing  sizes.  The  analysis  shows that these blocks can
be  moved  as  whole  entities  along ${\bar B_{2^n}}$ without
interference  from  each  other  resulting  in the exchange of
those  loops  along  which  they  move. The essential parts of
these  moves  can  be  represented  by  a  set  $A^{(n)}_m$ of
$2\pi$  movements  of  structural  blocks  $\Sigma^{(n)}_m$ so
that   $A^{(n)}_1$   corresponds  to  the  largest  block  and
results   in  an  exchange  involving  all  the  $2^n$  loops,
$A^{(n)}_2$  corresponds  to  movement  of  next-smaller braid
block  and  results in the exchange of $2^{n-1}$ loops, and so
on.  The  transformations  of time trajectories resulting from
exchanges  of  loops  can  be again described by the action of
permutation  operators  ${\cal A}^{(n)}_m$ on symbolic strings
$s_j$.  The  fact  that the crossing blocks move independently
results  in  the  commutivity  of operators ${\cal A}^{(n)}_m$
with  each  other.  The  geometry  of  ${\bar  B_{2^n}}$  also
defines  the  basic  algebraic  property of ${\cal A}^{(n)}_m$
demonstrated above for the $n=1,2$ examples
\begin{equation}
	({\cal A}^{(n)}_m)^{2^m} = {\bf 1}, \; m\in[1,n] .
\end{equation}	
Some    compositions   of   the   exchange   operators   yield
translation      operators      ${\cal      T}_{l}$      where
$l\in[-2^{n-1},2^{n-1})$.    For   the   discussion   of   the
phenomena  described  in  Sec.~\ref{local}  only  the operator
${\cal T}_{+1}$ and its inverse are of particular interest.

Using  induction  from  the  analysis  of cases with small $n$
one  may  infer the general expression for the ${\cal T}_{+1}$
translation operator :
\begin{equation}
\label{prod}
	{\cal T}_{+1} = \prod_{m=1}^n {\cal A}^{(n)}_m \;.
\end{equation}	

\end{appendix}

\end{multicols}


\begin{thebibliography}{99}
\bibitem{amend}  Since  in  the  following  we  consider local
trajectories  for  points  sufficiently far from a defect, the
difference between ${\bf c}^*_d$ and ${\bf c}^*$ is neglected.
\bibitem{f3}  The  difference between period 2 and period 4 is
not   clearly  seen  in  Fig.~\ref{4sp}  due  to  insufficient
spatial resolution of the phase using the color-coding scheme.
\bibitem{f4}  We  assume  that  the  oscillation  pattern  can
possess  period-4,  period-8  or  higher  order fine structure
but  we  are  only  concerned  with whether the representative
point falls within the smaller or larger band of period-2.
\bibitem{f5}  This  assumption is valid for the WR model since
the  motion  along  longer  loops  is  faster  than that along
shorter loops.
\bibitem{f9}  Since  operators  ${\cal  T}_{n_t}$  and  ${\cal
T}_{-n_t}$  are  equivalent  in  their  action  on  a period-2
oscillation  the  signs  are  chosen  to  preserve consistency
with cases $n>1$.
\end{thebibliography}
\end{document}